\documentclass[aps,reprint,showpacs,superscriptaddress,footinbib,citeautoscript]{revtex4-1}

\usepackage{graphicx}

\usepackage[utf8]{inputenc}
\usepackage{csquotes}
\usepackage[american]{babel}
\usepackage[T1]{fontenc}
\usepackage{enumerate}
\usepackage{mdwlist}
\usepackage[activate=normal]{pdfcprot}
\usepackage{bbding}
\usepackage{color}
\usepackage{amsmath}
\usepackage{eqnarray,amsmath}
\usepackage{amssymb}
\usepackage{amsmath}
\usepackage{amsfonts}
\usepackage{mathrsfs}
\usepackage[titletoc,title]{appendix}
\usepackage{bm}
\usepackage{dcolumn}
\usepackage{color}
\usepackage[colorlinks=true,citecolor=blue]{hyperref}
\hypersetup{colorlinks=true,citecolor=blue,linkcolor=red
,urlcolor=blue}

\begin{document}
\title{Spin transfer torque and exchange coupling in Josephson junctions with ferromagnetic superconductor reservoirs}
\author{Zahra Shomali}
\affiliation{School of Physics, Institute for Research in Fundamental Sciences (IPM), Tehran 19395-5531, Iran}
\author{Reza Asgari}
\affiliation{School of Physics, Institute for Research in Fundamental Sciences (IPM), Tehran 19395-5531, Iran}
\affiliation{School of Nano Science, Institute for Research in Fundamental Sciences (IPM), Tehran 19395-5531, Iran}
\date{\today}
\begin{abstract}
In this paper, the spin transfer torque (STT) and the exchange coupling of the Josephson junctions containing the interesting cases of diffusive/ballistic-triplet/singlet ferromagnetic superconductor (FS) materials together with the diffusive Josephson junction of the form S$_{1}$/F$_{1}$/I$_1$/N/I$_2$/F$_2$ with I being the insulating barrier are investigated. Using the Nazarov quantum circuit theory, it is found that for the diffusive FS$_1$/N/FS$_2$ structure the STT only appears in the normal direction to the plane of the exchange fields of the F$_1$ and F$_2$. This results in the antiparallel/parallel or vice versa parallel/antiparallel transition of the favorable exchange coupling depending on the considered parameters of the system, including the nonmagnetic spacer thickness, the superconducting phase difference, the length and the exchange field of the ferromagnets. Furthermore, the analyze of the width of the transitions, the phase difference interval in which an interlayer length-induced antiparallel/parallel transition can be occurred, is performed. For instance, as the exchange field or the temperature increases, the interval of phase difference gets larger. On the other hand, the ballistic Josephson junction containing the triplet ferromagnetic superconductor reservoirs solving the 16$\times$16 Bogoliubov-de-Gennes equation is studied. It is found that although the exchange fields of the FS are laid in the $z$ and $y$ direction, the STT interestingly exists in all three directions of $x$, $y$ and $z$. This exciting finding suggests that the favorable equilibrium configuration concerning the least exchange coupling occurs in the relative exchange field direction different from 0 or $\pi$. To the best of our knowledge it is for the first time that the occurrence of the in-plane STT is reported. Moreover, the occurrence of the beat like behavior with two oscillation period for the out-of-plane STT is interestingly acquired.
\end{abstract}
\maketitle
\section{\label{sec:intro}Introduction}
The potentiality of spin transfer torque for commercial application in MOSFET technology has been a strong driving force in this field from the beginning. Slonczewski's 1997 patent \cite{Slonczewski1997} for devices based on spin transfer torque(STT) has been referenced by 346 subsequent patents. Research on spin-transfer-induced domain wall motion has been highly pursued by many groups since the beginning of the 21st-century \cite{Aschenbach2000,Kasai2001,Cros2002,Bland2003,Parkin2003}. The coupling between itinerant carriers and collective magnetic order parameters implied in STT makes this quantity as the practical tool for the usage in magnetic random access memories and oscillator circuits \cite{Brataas2006,Slonczewski2011,Yulaev2011,Belanovsky2012,Locatelli2014,Apalkov2016}. In addition to extensive theoretical and experimental studies of the STT in ferromagnetic (F) spin-valve and domain structures, there have been studies devoted to the STT in structures with superconducting parts \cite{Waintal2002,Durst2005, Fazio2005,Escudero2007,Zhao2007,Zhao2008,Metalidis2009,Shomali2011,Shomali2011s,Muduli2017}. Actually, with novel and featuring applications, the ferromagnet superconductor heterostructures have been the active topics of the condensed-matter physics research \cite{Saxena2000,Aoki2001,Huy2007}. In particular, special attention has been paid to the Josephson nano systems including the FS structures. This is due to the spin sensitivity in ferromagnets in combination with the dissipation-less currents in superconductors, which open ways towards the novel types of controlled charge and spin flow nano machines. The created STT in these nanojunctions is the cause of the transfer of spin angular momentum from the spin current to the magnetization. The torque may have both in-plane and out-of-plane components. The equilibrium exchange interaction is brought on only an out-of-plane component, while the non-equilibrium torque is mainly in the plane. The equilibrium out-of-plane STT always tries to retain the minimum of the free energy at the state in which the exchange fields of the ferromagnets make the angle 0 or $\pi$ relative to the each other. While, the in-plane STT makes the magnetic moments switch to a perpendicular configuration, rather than a parallel or antiparallel one \cite{Waintal2000,Waintal2001s}. The sign and size of the equilibrium/non-equilibrium torque may be controlled by many parameters such as the direction of the spin supercurrent, the superconducting phase difference, the magnitude and the direction of the exchange couplings. Dealing with the formation of the STT in ferromagnetic superconductor heterostructures, the essential effect is the production of the long-ranged spin-triplet superconducting correlations by the interplay between the induced spin-singlet correlations and the noncollinearity of the magnetization profile in the F contact \cite{Bruno1995,Volkov2001,Volkov2001a,Volkov2001b,Blanter2004,Buzdin2005,Bergeret2005,Tollis2007,Braude2007,Braude2008,Alidoust2010}. Despite the aspects of the behavior of the STT in the ferromagnetic junctions including the s-wave superconductivity are known \cite{Waintal2002,Saidaoui2014,Halterman2015,Sirvasta2017,Halterman2018}, only one work has been devoted to the attitude of the torque in the junctions containing the triplet p-wave chiral ferromagnetic superconductor reservoirs. In the mentioned study, Linder \emph{et al.} \cite{Shomali2012} showed that the physics of the spin transfer torque appear richer and more complex in these structures. In particular, the STT depends on the phase of the superconducting pairing correlations which can be utilized as an additional way of controlling and detecting spin-transport and magnetization dynamics. On the other hand, it should be noted that only an out-of-plane component of the STT as the consequence of the equilibrium exchange interaction has been studied in this perusal. A chiral gap function breaks time reversal symmetry due to the appearance of the $i$ in the p-wave form of the paring potential, $\Delta\propto$(k${_x}\pm$ ik$_{y}$). Experimental confirmation of broken time-reversal and mirror symmetry of $^3$He-A was recently made by the observation of an anomalous Hall effect for electron bubbles moving in $^3$He-A films \cite{Ikegami2013,Wu2018}. We believe that the time-reversal breaking results in interesting findings for the spin transfer torques of these systems.

The indispensable need for the reduction of the bit size which consequently results in the smaller thinner electrical devices \cite{Shomali2016,Shomali2017,Shomali2018}, made the way toward the usage of the new propounded spintronics devices. The so-called spin MOSFET, encodes the binary information as spin. The MOSFET cell stores one bit of binary information through the parallel or antiparallel magnetization of the recommended MOSFET. In other words, the spin MOSFETs are designed to inaugurate two stable states of relative magnetization between the source and the drain ferromagnetism reservoirs, viz., parallel and antiparallel magnetization \cite{Atulasimha2016}. The STT is declared as an efficient tool to manipulate the magnetic states, the central information medium of spintronic devices, in an accurate, swift and at low energy cost way \cite{Locatelli2014,Linder2014,Baek2015}. For example, by applying a bias to the increasingly small magnetic MOSFET bits the readout operation is performed via the magnetoresistive effects just like what happens in the case of the flash memory. On the other hand, during the writing operation, the magnetization configuration of the spin MOSFET will be varied by some tools including the current induced magnetic field. There have been also proposals for the nonvolatile memory and the reconfigurable logic gates using spin MOSFETs, where the logic functions are modulated by reversing their magnetization configurations \cite{Tanaka2007}. In brief, the spin torque can be engineered to manufacture a various of progressive magnetic nano-devices. The conformation of the ferromagnets used in MOSFET qubits are usually being controlled by implying external magnetic fields so far. At the same time, these settled states with the minimum free energy configurations can be rather well regulated by finding more ways to change the favorable relative arrangements of the magnetization to mutate from the parallel design to antiparallel alignment and vice versa.

To get the full benefits of the potential of the ferromagnetic and superconductor structures, the Josephson junctions with diffusive magnetic junction and also with two FS reservoirs, connecting through the normal metal connector are the qualified candidates suggested as the novel spin MOSFET devices. In these structures the minimum free energy can be modulated by the new interventions such as, changing the relative superconducting phase differences $\phi$ of the two reservoirs, for the specified length of the normal layer. In other words, by varying $\phi$, one can manage the minimum of the free energy to occur in a parallel or an antiparallel configuration of the magnetic moments. In addition, the distinct magnetic couplings which lead to the parallel or the antiparallel alignment is favored by the reasons including the different thicknesses of the normal layer, the value of the exchange fields, the contemplated temperature, and so on, and so forth. However, further studies are required to fully understand the mechanism under which circumstance the preferable magnetic couplings occur.

In this paper, the STT and the exchange couplings in the recently recommended ferromagnetic Josephson junction MOSFET bits are investigated. Mainly, the standard equilibrium exchange interaction leads to the equilibrium Josephson-effect induced torque, which is perpendicular to the plane spanned by the directions of the magnetic moments of the ferromagnetic layers. On the contrary, the nonequilibrium torque typically stands inside the plane. The equilibrium exchange interaction results in the equilibrium spin current flowing from one ferromagnet to the other while the nonequilibrium torque levitates from the non-conservation of flowing spin currents in coexistence with the electrical current. At first, the diffusive ferromagnetic Josephson junctions are considered. More, two cases of S/F$_c$/S junction with the singlet superconductor reservoirs and the Josephson junction with the singlet ferromagnetic superconductor reservoirs, FS$_1$/F$_1$/N/F$_2$/FS$_2$, with specified temperature, exchange field, and the length of the ferromagnetic/normal layer will be explored. In the headmost studied structure, the F$_c$ represents a complex ferromagnetic junction of length L, which consists of diffusive ferromagnetic and normal metal parts as well as insulating barriers. In particular, the essential characteristic for the MOSFET qubit, the favorable configuration of the two ferromagnets relative to each other for the occurrence of the minimum of the free energy, is enquired. Considering the $\alpha$ as the angle between the exchange fields of the two FSs or the ferromagnets, the width of the phase difference in which the equilibrium configuration of $\alpha$=$\pi$ to $\alpha$=0 transition occurs in a diffusive Josephson junction will be also studied. We define the width of the transition as the interval of the superconducting phase difference, in which the transition from the antiparallel configuration to parallel or contrariwise is possible. It is obtained that only equilibrium STT appears in these diffusive Josephson junctions. Accordingly, the width of the transition of the favorable configuration of the exchange coupling from the antiparallel conformation to the parallel situation for four different structures with F$_c$ being F$_1$/N/F$_2$, symmetric F$_1$/I/N/I/F$_2$ and symmetric/asymmetric I$_1$/F$_1$/N/F$_2$/I$_2$ double barrier junctions is investigated. Here, "I" denotes the insulating barrier. For the S$_1$/I$_1$/F/I$_2$/S$_2$ structure we show that the existence of the I barriers at the SF interfaces broadens the 0 to $\pi$ transitions and, therefore, improves the conditions to easily trace the track of such transitions for the usage in MOSFET qubits. It will be demonstrated that a symmetric double-barrier structure with the two barriers having the same conductance shows wider transitions than the corresponding asymmetric structure with the same total conductance but different conductances of the barriers. An even larger width of transitions can be achieved by including an additional normal-metal part into the F$_c$. This motivates our study toward the symmetric S$_1$/I/F$_1$/N/F$_2$/I/S$_2$ Josephson structure, with the optimum relative width of $\phi$. Then, the condition in which the transition from ferromagnetic parallel configuration to antiparallel one for the diffusive singlet ferromagnetic superconductor Josephson junction occurs, are further scrutinized.

In the second part, the STT and the exchange coupling of the ballistic Josephson junction with triplet ferromagnetic superconductor reservoirs connecting through the short normal layer, FS$_1$/N/FS$_2$, are studied. The existence of the in-plane STT, which allows much more control on the intended magnetic nano-devices, is the reason for the importance of suchlike inquest. A thorough search of the relevant literatures yielded that the in-plane STT has not been investigated yet in these kinds of nano structures. As previously indicated, one of the fascinating properties of the STT is the ability to move the magnetization on the energy landscape. This behavior leads the magnetization to get the new equilibrium positions which do not belong to energy minima and it can establish the steady oscillations of the magnetization with large precession angles. The in-plane torque is modified when the F/N/F trilayer is coupled to the superconductor or FS reservoirs rather than connecting to two normal metal reservoirs. The superconductor reservoir contact conveys both spin and charge currents, while the normal one only sustains a charge current for voltages below the superconducting gap. It is perceived that this restriction causes a in-plane torque, which brings about the switching of the magnetic moments to a perpendicular configuration, rather than a parallel or antiparallel one \cite{Waintal2000,Waintal2001s}. The sign and size of the in-plane torque is controlled by the superconducting phase difference of the two FS reservoirs, the phase difference $\Delta$$\theta$=$\theta^{\uparrow}$-$\theta^{\downarrow}$ between the majority and minority spin superconducting order parameters, and so on.

In the following the structure of the paper is briefly described. First, in Sec. \ref{sec:2}, the model and the basic equations, used for investigation of the various diffusive/ballistic Josephson junctions are given. The obtained results will be discussed in Sec. \ref{sec:3}. Finally, the paper is concluded in Sec. \ref{sec:4}.

\section{Model and Basic Equations}
\label{sec:2}
In this Article, two different approaches are implied to deal with the two different cases of the diffusive and ballistic Josephson junctions. At first sight, the Usadel equation will be discretized by means of the Nazarov's quantum circuit theory for the diffusive nanosystems. Thereupon, the investigation of the ballistic Josephson structures is anteceded exploiting Bogoliubov-de-Gennes formalism (BDG). Two procedures will be more portrayed in the following two subsections.
\subsection{Diffusive MOSFET spin}
\begin{figure}
\vspace*{-2.8cm}
\centering
\hspace*{-0.35cm}
\vspace*{-2.9cm}
\includegraphics[scale=0.35]{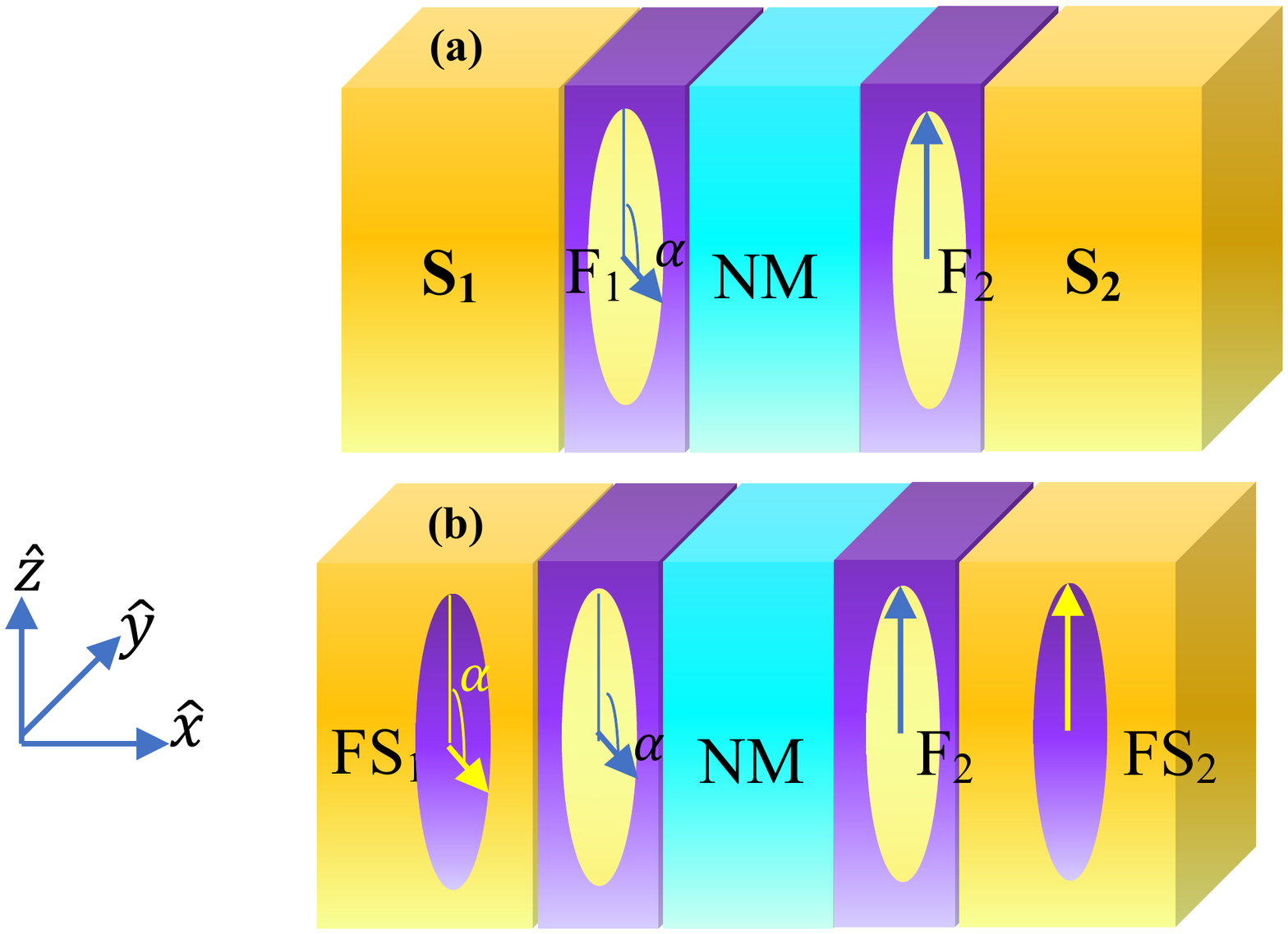}
\caption{\label{geometry1}(a) and (b) respectively demonstrate the studied geometry of S$_1$/F$_c$/S$_2$ with F$_c$ being F$_1$/N/F$_2$, symmetric F$_1$/I/N/I/F$_2$ and symmetric/asymmetric I$_1$/F$_1$/N/F$_2$/I$_2$ double barrier junctions and the FS$_1$/F$_1$/N/F$_2$/FS$_2$ structure.}
\end{figure}

The first part of the present study is devoted to the investigation of the minimum free energy established states of the diffusive S$_1$/F$_c$/S$_2$ Josephson junction with F$_c$ being different configurations of the ferromagnets, insulating barriers, and the superconductors and then the structure of the form FS$_1$/F$_1$/N/F$_2$/FS$_2$, exhibited in Figs. \ref{geometry1}(a) and (b).
In more details, primitively, the numerical calculations of the diffusive S$_1$/F$_1$/N/F$_2$/S$_2$, symmetric S$_1$/F$_1$/I/N/I/F$_2$/S$_2$, and symmetric/asymmetric S$_1$/I$_1$/F$_1$/N/F$_2$/I$_2$/S$_2$ double barrier junctions are performed, while at the second place the exchange coupling study of the diffusive FS$_1$/F$_1$/N/F$_2$/FS$_2$ Josephson structure is taken into account. FS$_1$ and FS$_2$ are, respectively, singlet ferromagnetic superconductor with the superconducting phase differences of $\phi_1$ and $\phi_2$ and the exchange field of h$_1$ and h$_2$ both in the $y$-$z$ plane. It is assumed that the exchange field of the FS$_1$ in similarity to that of F$_1$, makes an angle of $\alpha$ relative to the exchange field in the FS$_2$ in likeness to that of the F$_2$ in conventional ferromagnetic Josephson junction. The solving procedure is followed by discretizing the Usadel equation with the means of the so-called quantum circuit theory (QCT) which is a finite-element technique for dealing with the quasiclassical Green's functions in the diffusive limit. The QCT description has the advantage of not requiring specified the concrete geometry. By discretization of the Usadel equation, the relations in analog to the Kirchhoff laws of classical electric circuit theory are obtained. These equations will be solved numerically by the iterative methods and then consequently the quasiclassical Green's function of the whole system will be determined.

Moreover, the QCT has been generalized to spin-dependent transport in Refs. [\onlinecite{hernando1}] and [\onlinecite{hernando2}] where the findings of F/N contacts are extended to handle the case studies of S/F$_c$/S and FS/N/FS contacts. The F$_c$ interlayer between the superconducting reservoirs as well as the normal layer between the ferromagnetic superconductor reservoirs are discretized into nodes. As shown in Ref. [\onlinecite{hernando2}], the normalized density of states in the normal metal connected to a superconductor and an insulator ferromagnet at its ends, is the same as the one for a BCS superconductor in the presence of a spin-splitting magnetic field. Accordingly, every node in a ferromagnetic layer with the specific exchange field is equivalent to a normal node connected to a ferromagnetic insulator reservoir (FIR) with the same exchange field. On the other hand, the FS reservoirs are presented with the nodes attached to two both FIR and superconducting reservoirs while each S reservoir of S/F$_c$/S junction is displayed by a node connected only to the S reservoir. The conductances of the tunnel barriers at S/F$_1$, F/S$_2$, F/S$_1$, etc interfaces are shown respectively by g$_{S{_1}F}$, g$_{FS{_2}}$, g$_{FS_{2}F{_2}}$ and et cetera. Furthermore, the two nodes in the F$_c$ domains are assumed to be weakly coupled to each other by means of a tunneling contact. The conductance of such contact between each two nodes inside the F$_{c}$ is represented by g$_{T}$ determined by g$_{F_{c}}$, the total conductance of F$_{c}$ excluding the conductances of the barriers at the interfaces, ($n$-1)/g$_{T}$=(1/g$_{F_{c}}$)-(1/g$_{S_1F}$+1/g$_{FS_2})$. Here, $n$ is the total number of the contemplated nodes inside the F$_c$. In general, a node inside F$_c$ is characterized by a Green's function $\check{G}_{i}$, which is an energy-dependent 4$\times$4-matrix in the Nambu and spin spaces.  In terms of spin-space matrix components $\hat{g}$ and $\hat{f}$, the matrix Green's function is written as
\begin{equation}
  \check{G}=\left(
    \begin{array}{cc}
      \hat{g} & \hat{f} \\
      \hat{f}^{\dag} & -\hat{g}
    \end{array}
  \right),
  \hat{a}=\left(
    \begin{array}{cc}
      a_{\uparrow\uparrow} & a_{\uparrow\downarrow} \\
     a_{\downarrow\uparrow} & a_{\downarrow\downarrow}
    \end{array}\right),a=f,g.
\end{equation}
In the present case of study, a structure with spin-dependent magnetic contacts and in the presence of F and S reservoirs, the matrix current is developed in Ref. [\onlinecite{hernando2}]. The matrix current between two nodes named as $i$, $j$ has three components. The first term expressing the matrix currents from the neighboring nodes $i$-1 and $i$+1, which can be F, N, S or FS, is illustrated as $\check{I}_{ij}$=(g$_{ij}$/2)$\left[\check{G}_{i},\check{G}_{j}\right]$ with $[,]$ proffering the commutator
of the two involved parameters. The second one known as $\check{I}_{\omega i}$ is equal to -G$_{Q}$($\omega$/$\delta_{i}$)$\left[\check{\tau_{3}},\check{G}_{i}\right]$, with $\delta_i$ being the electronic level spacing of the node. While $\tau_{3}$ indicates the unit matrix in Nambu space. In QCT calculation, the dephasing of the electrons and holes due to their finite dwell time in the node $i$ is taken into account by the leakage matrix current which is proportional to the energy, $\epsilon$ replaced by the Matsubara frequency\cite{Nazarov1994,Nazarov1999,Nazarov2005}. The third one represents the leakage caused by the spin-splitting due to an exchange field $\vec{h}_i$ (G$_{Q}$$\equiv$e$^{2}$/2$\pi\hbar$ is the quantum conductance),
\begin{equation}
\label{eq:mxcurrent}
\check{I}_{si}=\frac{G_{MR}}{4}[\{
(\vec{h}_{i}.\vec{\hat{\sigma}})\hat{\tau}_{3},\check{G}_{i}\},\check{G}_{j}]
\end{equation}
\begin{equation}
\nonumber\\
 +[i\frac{G_{Q}}{\delta_{i}}(\vec{h}_{i}.\vec{\hat{\sigma}})\hat{\tau}_{3},\check{G}_{j}].
\end{equation}
In the above equation, $\{,\}$ presents the anticommutator of the matrices. Due to the relation G$_{MR}$$\sim$g$^{\uparrow}_{i,j}$-g$^{\downarrow}_{i,j}$$\ll$g$_{i,j}$, this term can be negligible. Besides, $\vec{h}$ is the exchange field of the node, $\vec{\sigma}$ and $\vec{\tau}$ are, respectively, the vectors consisting of Pauli matrices in spin and Nambu spaces. All these matrix currents entering one node should obey the following law of current conservation in the matrix form,
\begin{equation}
  \check I_{\omega i}+\check I_{s i}+\sum_{j=i\pm1}\check I_{ij}=0.
  \label{eq:balance}
\end{equation}
The Eq. (\ref{eq:balance}) is written for all nodes and then is supplemented by the boundary conditions, which are the values of $\check{G}$ in the S or FS reservoirs. Also, the inverse proximity effect in the reservoirs is neglected and the matrix Green's functions in S$_1$ and S$_2$ are set to the bulk values of
\begin{equation}
  \check{G}_{{1,2}}=\frac{\omega_{m}\check\tau_3+{\check\Delta}_{{1,2}}}{\sqrt{\omega_{m}^{2}+\left|\Delta\right|^2}},
\end{equation}
where
\begin{equation}
{\check{\Delta}}_{{1,2}}=\left(
         \begin{array}{cc}
           0 & {\hat\Delta}_{{1,2}} \\
           {\hat\Delta^{\dagger}}_{{1,2}} & 0 \\
         \end{array}
       \right)\,,\,
  \check\tau_3=\left(
    \begin{array}{cc}
      \hat{1} & \hat{0} \\
      \hat{0} & -\hat{1} \\
    \end{array}
  \right).
\end{equation}
Here, ${\hat\Delta}_{{1,2}}=|\Delta|\exp{(\pm i\varphi/2)}\hat\sigma_1$ are respectively the superconducting order parameter matrix in S$_1$ and S$_2$ or in FS$_1$ and FS$_2$ with $\hat\sigma_i$ being the Pauli matrices in spin space and $\omega_m$=(2m+1)$\pi$T with an integer number m and $k_B=1$, is the Matsubara frequency. The temperature dependence of the amplitude of the order parameter is taken into account as the well approximated value of $|\Delta|$=1.76T$_{c}$$\tanh$(1.74$\sqrt{T_c/T-1}$). Also, the matrix Green's function fulfills the normalization condition $\check{G}^{2}$=$\check{1}$. The size of F$_{c}$ is scaled in units of the diffusive superconducting coherence length, $\xi_{S}$=$\sqrt{\xi_{0}l_{imp}}$ with $\xi_{0}$=v$_{F}$/$\pi$$\Delta_{0}$ while v$_{F}$ is the Fermi velocity and $\Delta_{0}$=$\Delta$(T=0)=1.76T$_{c}$. Moreover, $l_{imp}$ is the mean free path in the F-layer related to the diffusion coefficient via D=v$^{(F)}_{F}$$l_{imp}$/3. Two more scales of h/T$_{c}$ and T/T$_{c}$ with T$_{c}$ being the critical temperature of the S or FS reservoirs, are further used. Also, the mean level spacing depends on the size of the system via the Thouless energy E$_{Th}$=D/L$^{2}$$\equiv$g$_{T}$$\delta$/($n$-1)G$_{Q}$ with $\hbar$=1. In the absence of spin-flip scatterings, the balance equation, Eq. (\ref{eq:balance}) is written for each spin direction separately for all the $n$ nodes in F$_{c}$ resulting in a set of equations for $n$ matrix Green's functions of the nodes which should be solved by numerical iteration methods. By choosing a trial form of the matrix Green's functions of the nodes for a given $\phi$, T, and the Matsubara frequency $m$=0 the calculation starts. Then, using Eq. (\ref{eq:balance}) and the boundary conditions iteratively, the initial value is refined until the Green's functions are computed with the desired accuracy for each of the $n$ nodes. The technique is fully described in the work by Shomali \emph{et al}. \cite{Shomali2008,Shomali2011,Shomali2011s}. Finding the Green's function of the systems, the components of the spin supercurrent and the STT are computed via the following formulas,
\begin{multline}
\label{Iij}
I=\textrm{tr}\hat\sigma_3\check I_{ij}, \ I_z=\textrm{tr}\check\tau_3\hat\sigma_3\check I_{ij}, \ I_{x(y)}=\textrm{tr}\hat\sigma_{1(2)}\check I_{ij};
\end{multline}

\begin{equation}
\nonumber
\check{I}_{ij}=(g_{ij}/2)\left [\check{G}_{i},\check{G}_{j}\right],
\end{equation}
and 
\begin{equation}
\tau_{zi}=I_{zi,i+1}+I_{zi,i-1}.
\end{equation}

As previously mentioned in Sec. \ref{sec:intro}, the aim of the first part of this paper is finding the equilibrium configuration of the exchange field vectors as a function of the phase difference and temperature to make them as the eligible candidate for the usage in MOSFET technology. Consequently obtaining the equilibrium STT, the exchange coupling will be appropriately calculated through the subsequent relation, 
\begin{equation}
F(\phi,\alpha)=\int_{0}^{\phi}I(\phi',\alpha)d\phi'+\int_{0}^{\alpha}I_z(\phi,\alpha')d\alpha'.
\end{equation}
The equilibrium angle, the angle between the two ferromagnets or the FSs in the equilibrium position, can be obtained by minimizing the free energy F, of the contact as a function of $\alpha$. In practice, it is found that the exchange field vectors favor either parallel, $\alpha$=0, or antiparallel $\alpha$=$\pi$ configurations. Pragmatically, the minimum of the free energy is calculated for the ferromagnetic Josephson junctions with very different structures and for various values of $\phi$, T/T$_c$, L$_N$/$\xi_s$, L$_F$/$\xi_s$ and h/T$_c$. Then for a specified structure the plots presenting L$_N$/$\xi_s$ versus $\phi$/$\pi$ are drawn distinguishing the region of the parallel equilibrium configuration from that of the antiparallel zone. Then the width of the transition is defined.  By the width we mean the interval of the superconducting phase difference, in which the transition from the antiparallel configuration to parallel or contrariwise is possible. As much as this width largens, the system affirms  better conditions for the transition from the parallel situation to the antiparallel one or in other words the detection of the parallel/antiparallel or vice versa transition can be more feasible for the structure having the larger width. The developed used code is obtainable at \onlinecite{Zahracode2}.
\subsection{Ballistic MOSFET spin}
\begin{figure}
\vspace*{-2.45cm}
\centering
\hspace*{-0.35cm}
\vspace*{-2.65cm}
\includegraphics[scale=0.25]{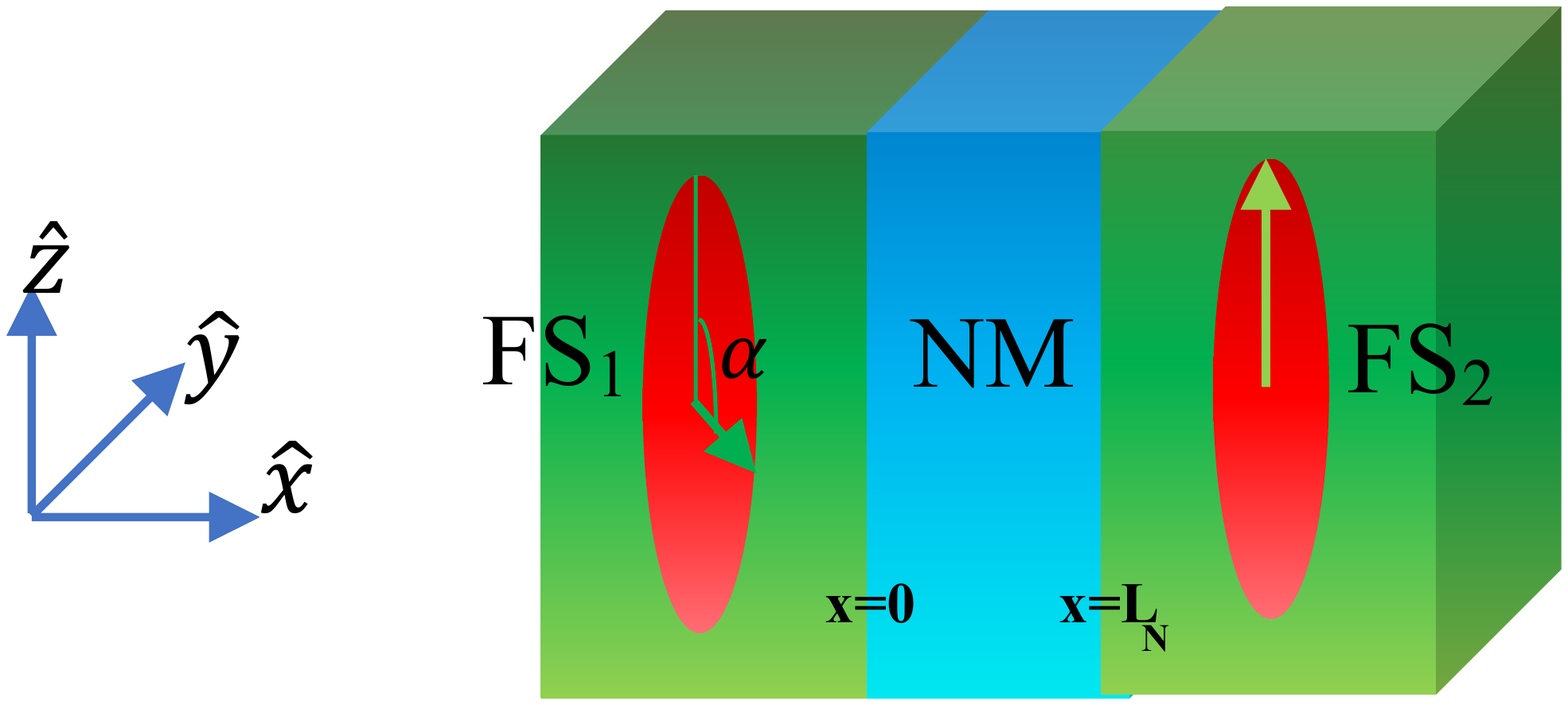}
\caption{\label{geometry2} The investigated ballistic geometry of FS$_1$/N/FS$_2$.}
\end{figure}

In the second part, the behavior of the STT in ballistic FS$_1$/N/FS$_2$ Josephson junction is taken into consideration. As shown in Fig. \ref{geometry2}, both FSs are two triplet FSs with the assigned phase differences of $\phi_1$ and $\phi_2$, the nonmagnetic spacer thickness, L$_N$, and the exchange fields of h$_1$ and h$_2$ making an angle $\alpha$ relative to each other. Further extra parameter of $\Delta \theta_{{1,2}}$=$\theta_{{1,2}}^{\uparrow}$-$\theta_{{1,2}}^{\downarrow}$, the phase difference between the majority and minority spin superconducting order parameters of the FS$_{1}$ and FS$_{2}$ is also contemplated. The quasiparticle propagation is described by the BDG equation,
\begin{equation*}
\hspace*{-0.8cm}\textbf{H}\psi(r)=E\psi(r)
\end{equation*}
for the four-vectors $\psi(r)=\left(
\begin{array}{cccc}
U_{\ell e(h)}^{\uparrow} & U_{\ell e(h)}^{\downarrow} & V_{\ell e(h)}^{\uparrow} & V_{\ell e(h)}^{\downarrow} \\
\end{array}
\right)
$, where the Hamiltonian can be written in the compact form as
\begin{equation}
\hspace*{-0.7cm} \textbf{H}=\left(
\begin{array}{cc}
H_{0}^{\ell} & \Delta_\ell \\
\Delta^{*}_\ell & -H_0^{\ell*} \\
\end{array}
\right)
\end{equation}
\begin{equation}
\hspace*{-0.7cm}
H_0^\ell=-\frac{\nabla^2}{2m}-E_F^{(\ell)}-{\bf{h}_{\ell}}\cdot{\bf{\sigma}},
\end{equation}
where, $\Delta_\ell$ is equal to $diag(\Delta_{\ell e(h)}^{\uparrow},\Delta_{\ell e(h)}^{\downarrow})$.
The wave functions in the three regions are presented in Appendix A; Eq. (\ref{1}-\ref{3}). Also the boundary conditions of Eq. (\ref{boundary}) will be applied. Then equivalent of the equations, the 16$\times$16 complex arbitrary matrix of the form \ref{4} are acquired. The STT will be then calculated by finding both the eigen values and eigen vectors of the desired matrix by using Lapack library \cite{Lapack1999} through the relation
 \begin{equation}
\label{tausst}
\hspace*{-0.3cm}\tau_{stt}=\{ \psi^{\dag}\left(
\begin{array}{cc}
\sigma\times \textbf{h} & 0 \\
0 & -\sigma^{*}\times \textbf{h} \\
\end{array}
\right)
\psi \},
\end{equation}
with
\begin{equation}
\hspace*{-0.8cm} \left(
\begin{array}{cc}
\sigma & 0 \\
0 & \sigma^* \\
\end{array}
\right)=\left(
\begin{array}{cccc}
\hat{z} & \hat{x}-i \hat{y} & 0 & 0 \\
\hat{x}+i \hat{y} & -\hat{z} & 0 & 0 \\
0 & 0 & \hat{z} & \hat{x}+i \hat{y} \\
0 & 0 & \hat{x}-i \hat{y} & -\hat{z} \\.
\end{array}
\right)
\end{equation}
Using the Eq. \ref{tausst}, the following spin transfer torques are obtained,
\begin{multline}
\hspace*{-0.43cm}
\tau_x=\frac{h}{E_F}(\sin\alpha\{-(|t_{e}^{\downarrow}|^{2}+|t_{h}^{\downarrow}|^{2})(|u_{2e}^{\downarrow}|^{2}+|u_{2e} ^{\downarrow}|^{2})+(|t_{e}^{\uparrow}|^{2}+|t_{h}^{\uparrow}|^{2}) \\ (|u_{2e}^{\uparrow}|^{2}+|u_{2e} ^{\uparrow}|^{2})+2\textrm{Re}\{-(v_{2e}^{\downarrow\star}u_{2e}^{\downarrow})t_{e}^{\downarrow\star}t_{h}^{\downarrow}e^{-i(q_{2e}^{\downarrow}+q_{2h}^{\downarrow})L}+\\(u_{2e}^{\uparrow\star}v_{2e}^{\uparrow}e^{-\bar{\gamma}_{2+}^{\uparrow}}+v_{2e}^{\uparrow\star}u_{2e}^{\uparrow}e^{\bar{\gamma}_{2+}^{\uparrow}})t_{e}^{\uparrow\star}t_{h}^{\uparrow}e^{-i(q_{2h}^{\uparrow}-q_{2e}^{\uparrow})L}\}\}+2\cos\alpha \\ \{\textrm{Im}\{v_{2e}^{\downarrow\star}(v_{2e}^{\uparrow}t_{h}^{\downarrow\star}t_{h}^{\uparrow}e^{-\bar{\gamma}_{2+}^{\uparrow}}e^{-\bar{\gamma}_{2-}^{\downarrow}}e^{i(q_{2h}^{\downarrow}-q_{2h}^{\uparrow})L})+u_{2e}^{\downarrow\star}(u_{2e}^{\uparrow}t_{e}^{\downarrow\star}t_{e}^{\uparrow}\\e^{i(q_{2e}^{\uparrow}-q_{2e}^{\downarrow})L}+v_{2e}^{\uparrow}t_{e}^{\downarrow\star}t_{h}^{\uparrow}e^{-\gamma_{2-}^{\uparrow}}e^{-i(q_{2e}^{\downarrow}+q_{2h}^{\uparrow})L})+v_{2e}^{\uparrow\star}(v_{2e}^{\downarrow}t_{e}^{\uparrow\star}t_{e}^{\downarrow}e^{\gamma_{2+}^{\uparrow}}\\e^{-\gamma_{2+}^{\downarrow}}e^{i(q_{2e}^{\downarrow}+q_{2e}^{\uparrow})L}+u_{2e}^{\downarrow}t_{e}^{\uparrow\star}t_{h}^{\downarrow}e^{\gamma_{2-}^{\uparrow}}e^{-i(q_{2e}^{\uparrow}+q_{2h}^{\downarrow})L})+u_{2e}^{\uparrow\star}(v_{2e}^{\downarrow}t_{h}^{\uparrow\star}t_{e}^{\downarrow}\\e^{-\gamma_{2+}^{\downarrow}}e^{i(q_{2h}^{\uparrow}+q_{2e}^{\downarrow})L}+u_{2e}^{\downarrow}|t_{2h}^{\downarrow}|^{2}e^{i(q_{2h}^{\uparrow}-q_{2h}^{\downarrow})L})\}\}),
\end{multline}
\begin{multline}
\hspace{-0.43cm}
\tau_y=\frac{h}{E_{F}}\sin\alpha(2\textrm{Re}\{t_{e}^{\uparrow\star}t_{e}^{\downarrow}e^{i(q_{2e}^{\downarrow}-q_{2e}^{\uparrow})L}((v_{2e}^{\uparrow\star}v_{2e}^{\downarrow}e^{\gamma_{2+}^{\uparrow}}e^{-\gamma_{2+}^{\downarrow}})-\\ (u_{2e}^{\uparrow\star}u_{2e}^{\downarrow}))+t_{h}^{\uparrow\star}t_{e}^{\downarrow}e^{i(q_{2h}^{\uparrow}+q_{2e}^{\downarrow})L}(u_{2e}^{\uparrow\star}v_{2e}^{\downarrow}e^{-\gamma_{2+}^{\downarrow}})-v_{2e}^{\uparrow\star}u_{2e}^{\downarrow}e^{-\bar{\gamma}_{2-}^{\uparrow}})\\+
t_{e}^{\uparrow\star}t_{h}^{\downarrow}e^{-i(q_{2e}^{\uparrow}+q_{2h}^{\downarrow})L}(v_{2e}^{\uparrow\star}
u_{2e}^{\downarrow}e^{\gamma_{2+}^{\uparrow}}-u_{2e}^{\uparrow\star}v_{2e}^{\downarrow}e^{\bar{\gamma}_{2-}^{\downarrow}})+t_{h}^{\uparrow\star}\\t_{h}^{\downarrow}e^{i(q_{2h}^{\uparrow}-q_{2h}^{\downarrow})L}(u_{2e}^{\uparrow\star}u_{2e}^{\downarrow}-v_{2e}^{\uparrow\star}v_{2e}^{\downarrow}e^{\bar{\gamma}_{2-}^{\downarrow}}e^{-\bar{\gamma}_{2-}^{\uparrow}})\}),
\end{multline}
and
\begin{multline}
\hspace{-0.43cm}
\tau_z=\frac{h}{E_{F}}\cos\alpha(2\textrm{Re}\{t_{e}^{\uparrow\star}t_{e}^{\downarrow}e^{i(q_{2e}^{\downarrow}-q_{2e}^{\uparrow})L}(u_{2e}^{\uparrow\star}u_{2e}^{\downarrow}-v_{2e}^{\uparrow\star}v_{2e}^{\downarrow}\\e^{\gamma_{2+}^{\uparrow}}e^{\gamma_{2-}^{\downarrow}})+t_{e}^{\uparrow\star}t_{h}^{\downarrow}e^{-i(q_{2e}^{\uparrow}+q_{2h}^{\downarrow})L}(u_{2e}^{\uparrow\star}v_{2e}^{\downarrow}e^{-\bar{\gamma}_{2+}^{\downarrow}}-v_{2e}^{\uparrow\star}u_{2e}^{\downarrow}e^{\gamma_{2+}^{\uparrow}})\\+t_{h}^{\uparrow\star}t_{e}^{\downarrow}e^{i(q_{2h}^{\uparrow}+q_{2e}^{\downarrow})L}(v_{2e}^{\uparrow\star}u_{2e}^{\downarrow}e^{-\bar{\gamma}_{2-}^{\uparrow}}-u_{2e}^{\uparrow\star}v_{2e}^{\downarrow}e^{\gamma_{2-}^{\downarrow}})+t_{h}^{\uparrow\star}t_{h}^{\downarrow}\\e^{i(q_{2h}^{\uparrow}-q_{2h}^{\downarrow})L}(v_{2e}^{\uparrow\star}v_{2e}^{\downarrow}e^{-\bar{\gamma}_{2+}^{\downarrow}}e^{-\bar{\gamma}_{2-}^{\uparrow}}-u_{2e}^{\uparrow\star}u_{2e}^{\downarrow})\}).
\end{multline}
The parameters $q_{\ell e(h)}^\sigma$  and $v(u)_{2e}^{\uparrow(\downarrow)}$ are described in Eqs. \ref{A4} and \ref{eq.v} of Appendix. Also, we define $r(t)_{e(h)}^{\uparrow(\downarrow)}$ in Appendix. Further, the used parameters are scaled in the following way
\begin{equation}
\hspace*{-0.8cm} \frac{k_x^\pm}{k_F}=\sqrt{(1 \pm \frac{E}{E_F})-\frac{k_y}{k_F}^2}
\end{equation}
and
\begin{multline}
\hspace*{-0.4cm} \frac{q_{1e(h)}^{\uparrow(\downarrow)}}{k_F}=\\\sqrt{1+\sigma \frac{h_1}{E_F} \pm \sqrt{\frac{E}{E_F}^2-(\Delta_{1\uparrow(1\downarrow)}/E_F )^2}-\frac{k_y}{k_F}^2}.
\end{multline}
Hence, the applied parameters are E/E$_F$, h$_{{1,2}}$/E$_F$, $k_y$/$k_F$ and consequently the length of the normal layer, L, is scaled to the fermi-wavelength $\lambda_F$. On the other hand the superconducting coherence length, $\xi_s$ is related to the Fermi velocity by the formula $\xi_s$=$\frac{\hbar v_{F}}{\pi\Delta}$. Recalling $v_F$ as $\frac{p_{F}}{m}$ and p$_{F}$ being $\sqrt{2mE_{F}}$, the $\xi_s$/$\lambda_F$ will be acquired as 1/$\pi^{2}(\Delta$/E$_F$). Then the length can be also scaled to the coherence length vis $\frac{L}{\xi_s}$=$\frac{L}{\xi_F}$$(\frac{\pi^2}{\Delta/E_F})$. Accordingly, the STT is scaled to $\sqrt{\frac{2E_F}{m}}$. The complete and precise procedure of the calculation is available in Appendix A. For subgap energies E$\leq|\Delta_{\uparrow,\downarrow}|$, successive Andreev Reflection (AR) at the two N/FS interfaces and the coherent propagation of the excitations between these reactions lead to the formation of so-called Andreev bound states (ABSs). In our FS$_1$/N/FS$_2$ system, these are correlated electron-hole pairs in spin-triplet states with a non-collinear polarization. The spectrum of the ABSs is obtained and then the calculated bound states are set in the formulations to solve the matrix equation. The recent experiment presents the evidence for STT induced by the spin triplet supercurrents where the resonance field is found to shift rapidly to a lower field below superconducting transition temperature T$_c$. They have reported the appearance of the STT when the thickness of the magnetic layer say is usually below the two nm \cite{Li2018}. Also, here the study is restricted to the most relevant case of a short N contact with the thickness L that is much smaller than the superconducting coherence length $\xi$. In this limit, the subgap Andreev states with E$\leq |\Delta_{\uparrow,\downarrow}|$ give the main contribution to the superconducting transport properties in the FS$_1$/N/FS$2$ structure and the contribution of the states of E$>|\Delta_{\uparrow,\downarrow}|$ can be disregarded. The opening of a gap at the Fermi level removes part of the normal-state exchange torque, and it is contemplated that for the E$\leq |\Delta_{\uparrow,\downarrow}|$ Andreev-states dominate the net exchange interaction.
\section{Discussion and Results}
\label{sec:3}
In the following subsection, foremost the results of the analysis of the exchange coupling for various diffusive singlet Josephson junctions are depicted. Afterwards, the outcomes pertinent to the ballistic triplet Josephson junctions will be given out.
\subsection{Diffusive Josephson junction}
This section imprimis presents the verification for the accuracy of the exploited method, then it proclaims the findings for the exchange coupling behavior of the conventional diffusive singlet ferromagnetic Josephson junction with miscellaneous barriers. Accordingly, the section winds up with the results obtained for the Josephson junction with singlet FS reservoirs.
\subsubsection{Verification}
\begin{figure}
\vspace*{-0.05cm}
\centering
\hspace*{-0.45cm}
\vspace*{-0.0cm}
\includegraphics[scale=0.61]{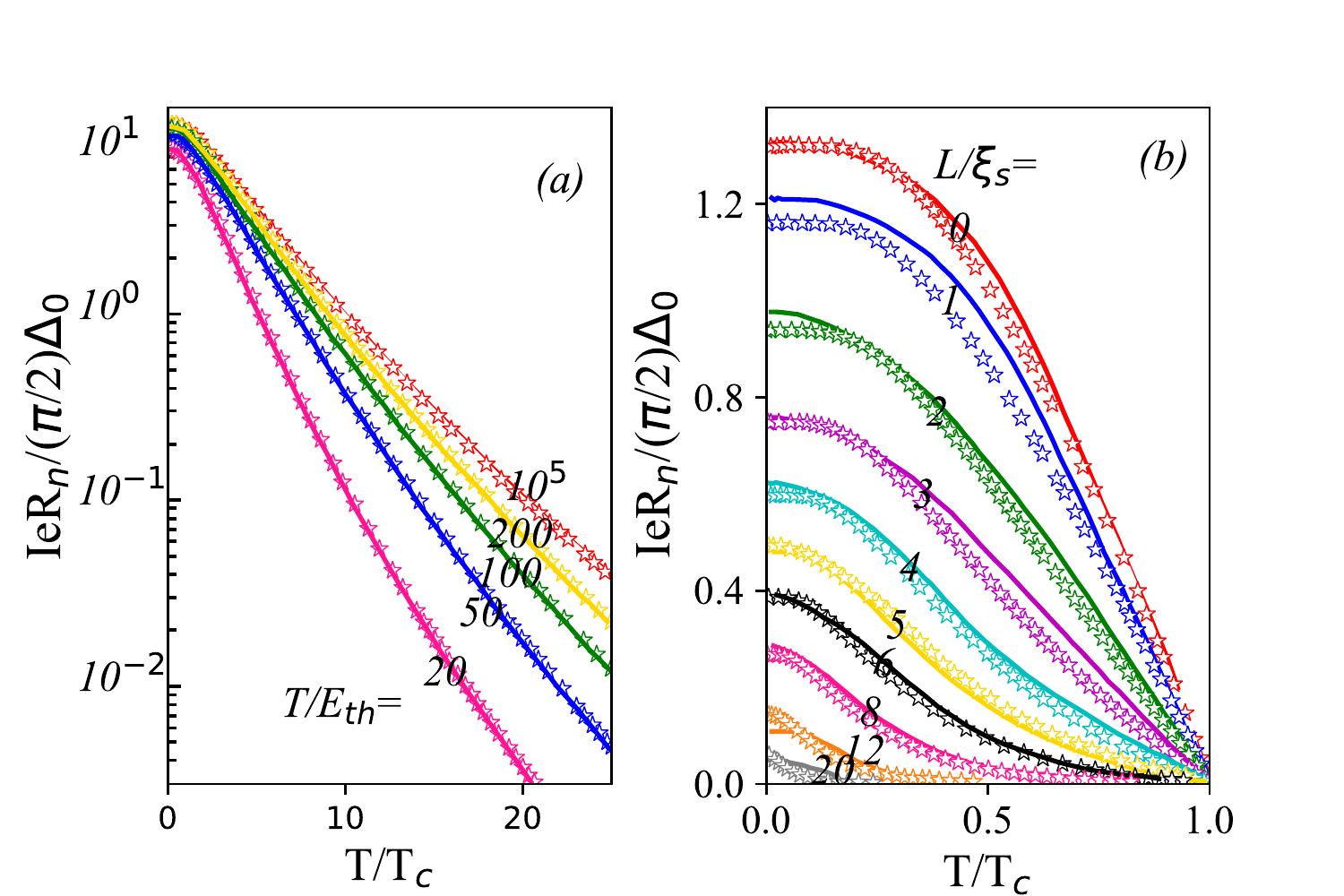}
\caption{\label{likharev2}(a) The presented study plot for different values of
$\frac{\Delta}{E_{Th}}$ (the star markers *) and the plot from Dobus's experimental work \cite{Dobus2001} (solid lines). (b) The critical current for different values of $\frac{L}{\xi_s(T_{c})}$ obtained from the present study (the star markers *) and the data extracted from the work by Likharev \cite{Likharev1979} (solid lines). }
\end{figure}

The method is verified by reproducing the experimental results presented by Dobus \emph{et al.} and the finding of Likharev for the behavior of the critical current of the S/N/S Josephson junction. Firstly, in similarity to the Dobus's work \cite{Dobus2001} the logarithmic plot for critical current versus temperature is drawn in Fig. \ref{likharev2}(a) when $\frac{T}{E_{th}}$=5.7$\times$10$^{-3}$ and $\frac{\Delta}{E_{th}}$=10$^{-4}$. As it is obvious our results shown by the star markers fit very well to the experimental outcomes of Dobus \cite{Dobus2001} manifested by solid lines. At the next step, the written code through QCT method is checked by comparing the obtained results with that of the Likharev \cite{Likharev1979}. The results are plotted for different values of $\frac{T_{c}}{E_{th}}$ but the scaling used in Likharev was $\frac{L}{\xi_s(T_{c})}$. These two diagrams are related to each other by the below formula
\begin{equation}
\xi_s(T_{c})=\sqrt{\frac{D}{2\pi T_{c}}}\nonumber\\
\end{equation}
\begin{equation}
(\frac{L}{\xi_s(T_{c})})^{2}=\frac{2\pi T_{c}L^{2}}{D}\nonumber\\
\end{equation}
\begin{equation}
E_{th}=\frac{D}{L^{2}}\nonumber\\
\end{equation}
\begin{equation}
\Rightarrow(\frac{L}{\xi_s(T_{c})})^{2}=2\pi \frac{T_{c}}{E_{th}}
\end{equation}
Fig. \ref{likharev2}(b) demonstrates the critical current versus t=T/T$_{c}$ for different values of $\frac{L}{\xi}$. As the Figs. \ref{likharev2}(b) show there exist a good consistency with the results obtained by the QCT method presented by star markers, and the previously experimentally and computationally acquired findings displayed by solid lines. The written code for the mentioned verification is procurable at \onlinecite{Zahracode1}.
\subsubsection{Symmetric Josephson junctions}
\begin{figure}
\vspace*{-0.7cm}
\centering
\hspace*{-0.52cm}
\vspace*{-0.7cm}
\includegraphics[scale=0.62]{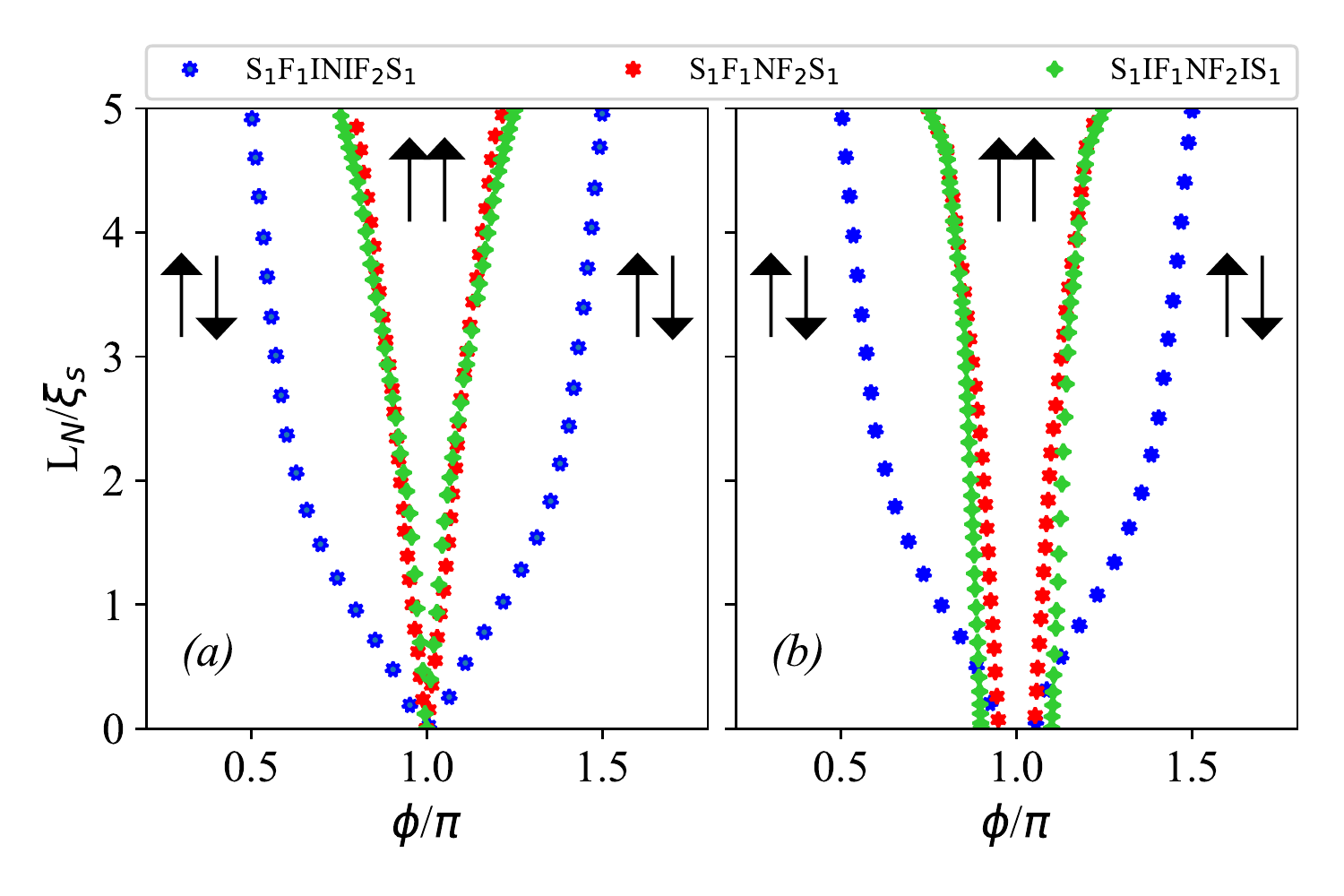}
\caption{\label{fig2}(a) Phase diagram of the favorable configuration of the junction versus $\phi$/$\pi$ for different types of structures when h/T$_c$=10, L$_F$/$\xi_s$=0.1, and, T/T$_c$=0.1 displaying the S$_1$/F$_1$/I/N/I/F$_2$/S$_2$ as the one with the largest value of $\Delta\phi$/$\phi_i$. Increasing the phase difference upto the value of $\phi$=$\pi$, the beginning of the normal layer induced transition decreases toward the smaller length of $L_N$/$\xi_s$=0.1 (b) The same as (a) but for h/T$_c$=100 resulting in bigger $\Delta\phi$/$\phi_i$ for all three symmetric structures. Note that the ratio $\Delta \phi$/$\phi_i$ becomes smaller for all three symmetric structures while the phase difference in which the system starts to prefers parallel configuration  decreases, whatever the length of the normal interlayer is.}
\end{figure}

At first place, the symmetric Josephson structures of S$_1$/F$_1$/N/F$_2$/S$_2$, S$_1$/I/F$_1$/N/F$_2$/I/S$_2$, and S$_1$/F$_1$/I/N/I/F$_2$/S$_2$ are studied. Symmetric structure is a contact with symmetric barriers satisfying the relation g$_{S_{1}F_{1}}$=g$_{S_{2}F_{2}}$. The Fig. \ref{fig2}, presents the transition of the favorable configuration with least energy of such nanosystems from antiparallel to parallel transition for the structures with specification of L$_F$/$\xi_s$=0.1, h/T$_c$=10, and T/T$_c$=0.1. Increasing the phase difference $\phi_i$ upto the value of $\phi$=$\pi$, the beginning of the normal layer induced transition decreases toward the smaller length of L$_N$/$\xi_s$=0.1. As seen in Fig. \ref{fig2}(a), it is obtained while the width of the transition $\Delta\phi$/$\phi_i$ is nearly the same for S$_1$/F$_1$/N/F$_2$/S$_2$ and S$_1$/I/F$_1$/N/F$_2$/I/S$_2$ structures, it increases notably for the symmetric junction of the form S$_1$/F$_1$/I/N/I/F$_2$/S$_2$. Also in this study, the effect of the exchange fields of the magnetic parts on the behavior of the transitional width is investigated. Figure. \ref{fig2}(b) manifests the phase diagrams for the system with the exchange field of h/T$_c$=100. It is acquired that the ratio $\Delta \phi$/$\phi_i$ becomes smaller for all three symmetric structures while also the phase difference in which the system starts to prefers parallel configuration, whatever the length of the normal interlayer is, decreases. Moreover, it is shown that the type of the transition strongly depends on the place of the barriers in such S$_1$/F$_1$/N/F$_2$/S$_2$ structures when L$_F$/$\xi_s$ is equal or larger than 0.1. Specifically, it is found while for the systems with symmetric barriers, the antiparallel-parallel transition occurs, the structure with no barrier shows parallel-antiparallel transition depending on the phase difference between the superconductor $S_1$ and $S_2$.
 
\begin{figure}
\vspace*{-0.7cm}
\centering
\hspace*{-0.29cm}
\vspace*{-0.2cm}
\includegraphics[scale=0.62]{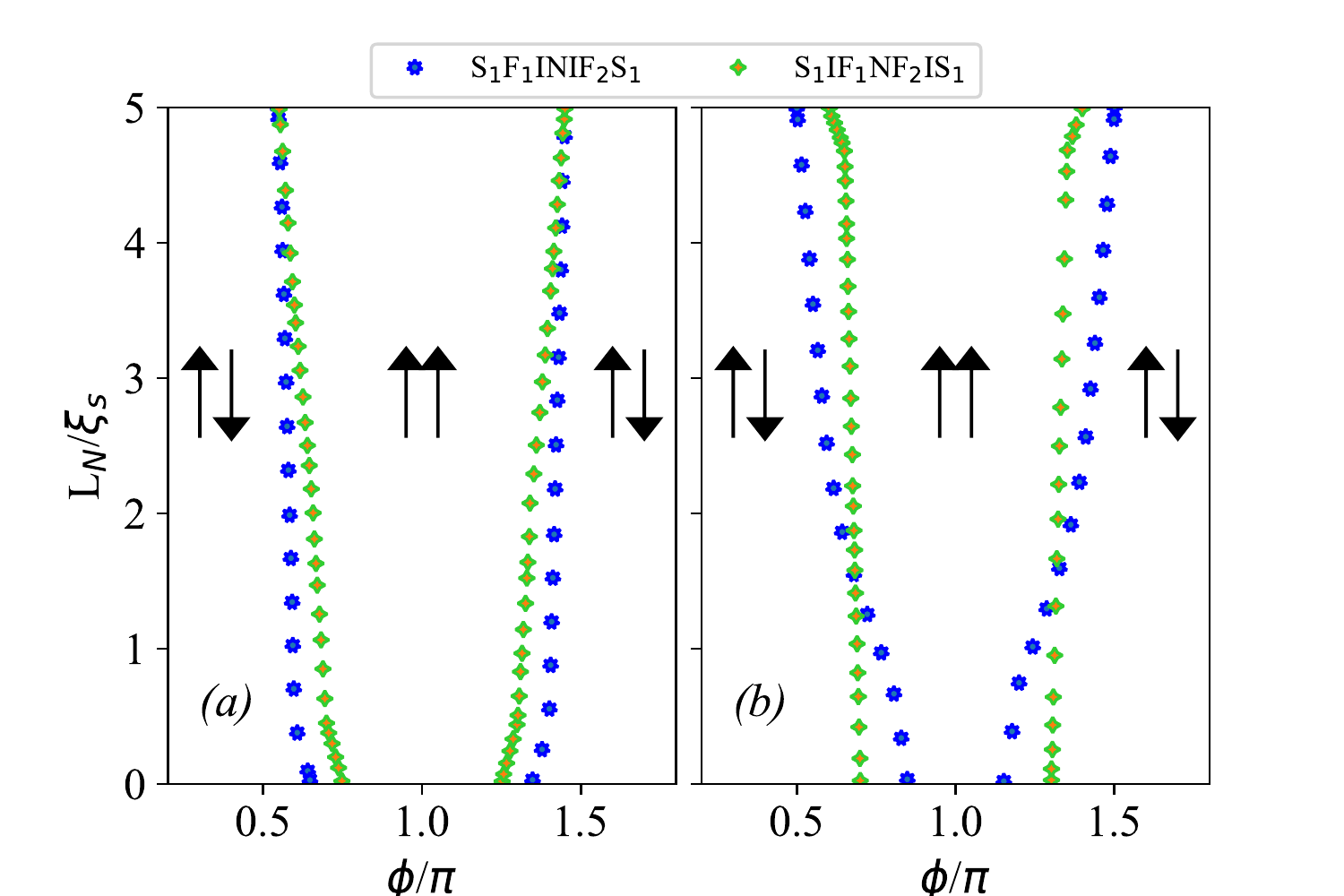}
\caption{\label{fig3} The same as Fig. \ref{fig2}(a) but for L$_F$/$\xi_s$=0.5. (b) Same as (a) but for T /T$_c$=0.5. Temperature increase evinces the change of the trend of the width $\Delta\phi$/$\phi_i$ for all three studied symmetric structures. The symmetric barrier systems can display the broader transition interval depending on the scale temperature T/T$_c$.}
\end{figure}

The Figs. \ref{fig3}(a) and (b) represent the exchange coupling behavior, respectively, for the structures with symmetric barriers of the form S$_1$/F$_1$/I/N/I/F$_2$/S$_2$ and S$_1$/I/F$_1$/N/F$_2$/I/S$_2$. It is figured out that depending on the scale temperature T/T$_c$, any of the symmetric barrier systems can display the broader transition interval. At low temperatures, the S$_1$/F$_1$/I/N/I/F$_2$/S$_2$ structure present wider interval of parallel-parallel equilibrium configuration of the exchange field vectors. In other words, minimizing the free energy F of the contact as a function of $\alpha$ results in $\alpha$=0 for more values of $\phi$ and L$_N$/$\xi_s$. This behavior of the favorable configuration of the desired Josephson systems reverses at high temperatures.

\begin{figure}
\vspace*{-0.4cm}
\centering
\hspace*{-0.4cm}
\vspace*{-0.2cm}
\includegraphics[scale=0.63]{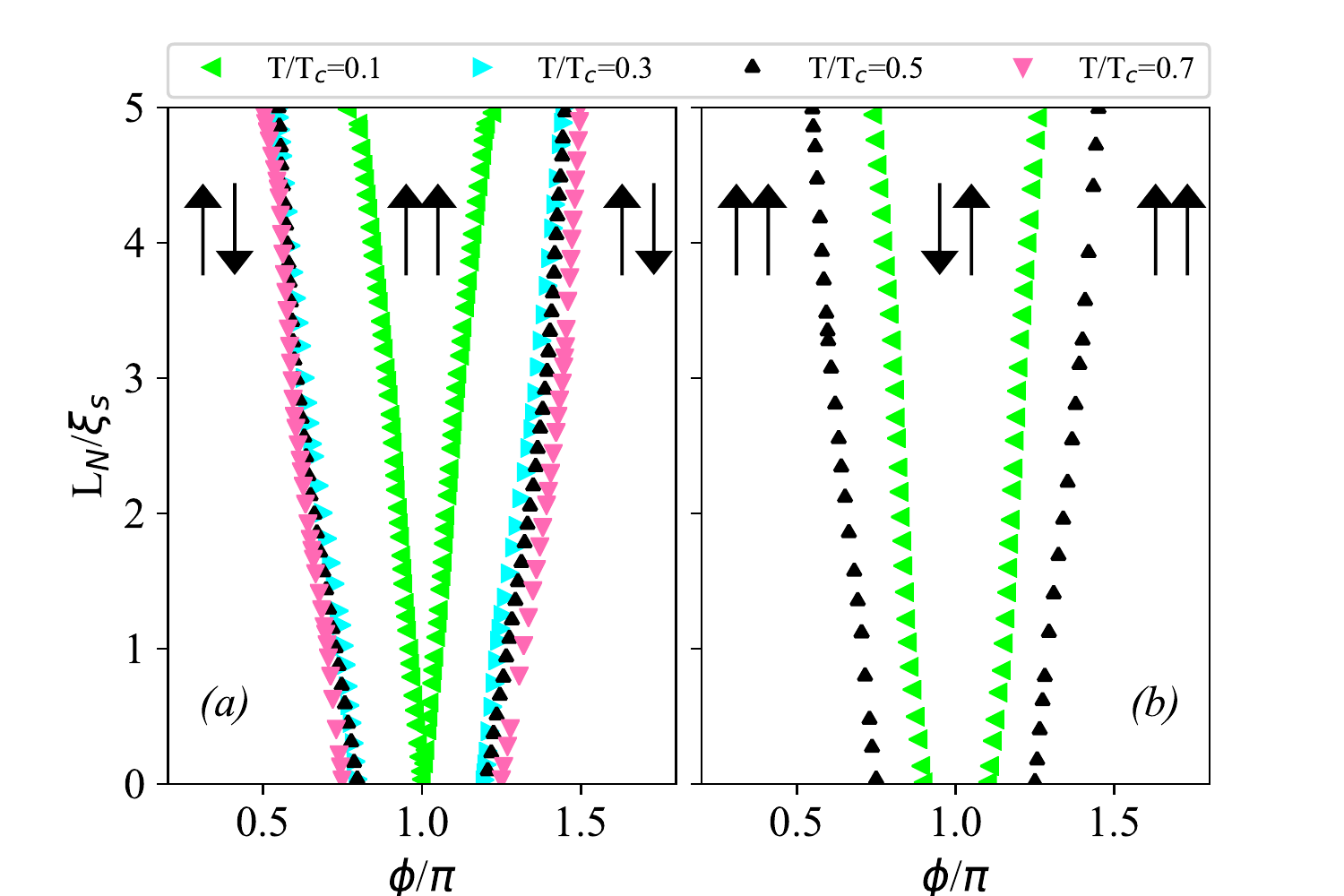}
\caption{\label{fig4} Phase diagram of the favorable configuration of the junction versus $\phi$/$\pi$ for S$_1$/I/F$_1$/N/F$_2$/I/S$_2$ structure when h/T$_c$=10 and L$_F$/$\xi_s$=0.1 for different values of temperature. Temperature increasing makes the antiparallel-parallel transition occur for smaller phase differences (b) The same as (a) but for S$_1$/F$_1$/N/F$_2$/S$_2$ structure presenting that the temperature increase causes the antiparallel-parallel transition happening for smaller phase differences.}
\end{figure}

Put differently, the effect of the temperature on exchange coupling is investigated. As presented in Fig. \ref{fig4}(a), for the system S$_1$/I/F$_1$/N/F$_2$/I/S$_2$, while keeping the exchange field at h/T$_c$=10 and the ferromagnetic length at L$_F$/$\xi_s$=0.1, temperature increasing makes the antiparallel-parallel transition occur for smaller phase differences. Further nano systems with higher temperature, experience much larger interval of antiparallel to parallel transition. The same happens as the L$_F$/$\xi_s$ largens which eventuates in augmentation of the interval of the occurrence of the antiparallel-parallel transition.
On the other hand, Fig. \ref{fig4}(b) exhibits the behavior of the minimum of the exchange energy for the system presenting no barrier, S$_1$/F$_1$/N/F$_2$/S$_2$, through the structure. It is shown that unlike what already obtained for the previous systems, at first place for low phase differences, the equilibrium configuration of the exchange vectors shows parallel-parallel configuration. Increasing the phase difference, $\Delta \phi$, at the specified phase of $\phi_i$ the system starts to experience the antiparallel-parallel equilibrium formation.

\begin{figure}
\vspace*{-0.0cm}
\centering
\hspace*{-0.2cm}
\vspace*{-0.0cm}
\includegraphics[scale=0.6]{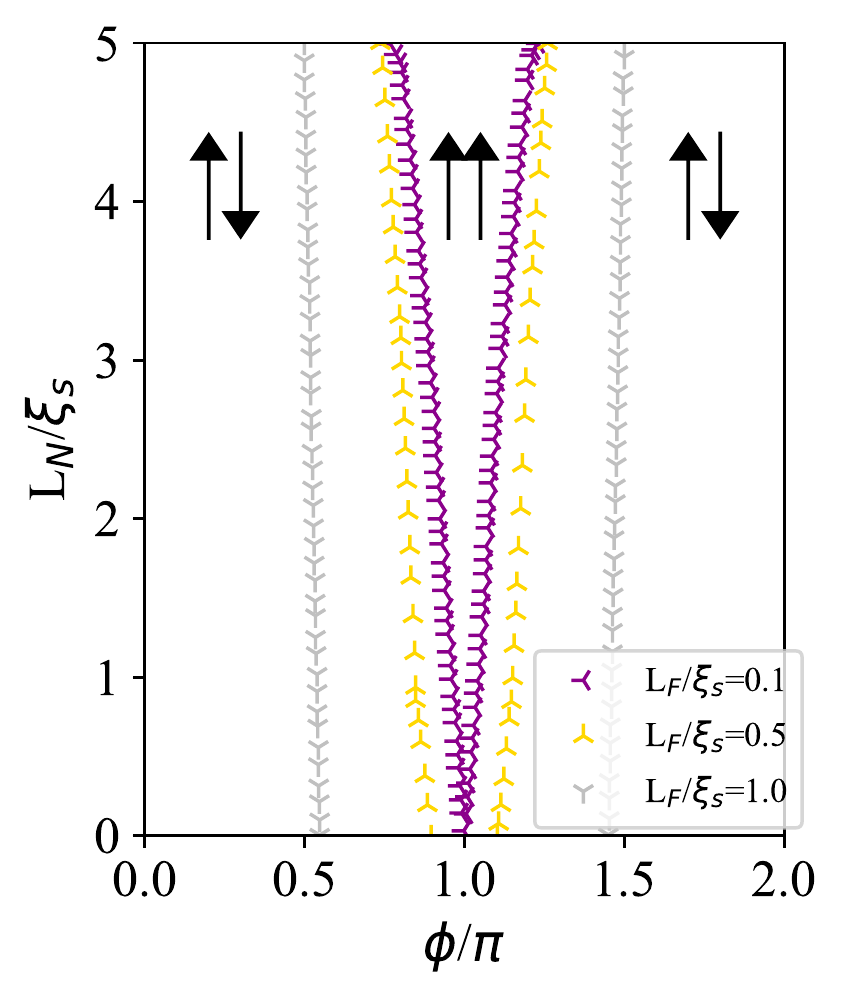}
\caption{\label{fig6} Phase diagram of the favorable configuration of the junction versus $\phi$/$\pi$ for S$_1$/F$_1$/N/F$_2$/S$_2$ structure with different L$_F$/$\xi_s$ lengths when h/T$_c$=10 and T/T$_c$=0.1 illustrating the growth of $\Delta \phi$ by the ferromagnetic length increase.}
\end{figure}

Temperature dependency of the favorable equilibrium configuration of S$_1$/F$_1$/N/F$_2$/S$_2$ structure is similar to that of the S$_1$/I/F$_1$/N/F$_2$/I/S$_2$. Moreover, the length of the ferromagnetic layer has been studied as the variable parameter. Specifically, Fig. \ref{fig6} manifests the outcome of such detection. Increasing the value of L$_F$/$\xi_s$ from 0.1 to 1.0, the $\Delta \phi$/$\phi_i$ becomes notably large, insofar as it reaches a saturation point of 1.64. In other words the $\Delta \phi$/$\pi$ interval in which the antiparallel-parallel transition occurs, grows from zero for L$_F/\xi_s$=0.1 upto 0.9 for the length L$_F/\xi_s$=1.0, while being independence to the value of the normal length. Put differently, the effect of increasing the ferromagnetic length layer on the antiparallel-parallel transition is similar to that of the enhancement of the exchange field for all three studied symmetric structures.
\subsubsection{Asymmetric Josephson junctions}
\begin{figure}
\vspace*{-0.0cm}
\centering
\hspace*{-0.4cm}
\vspace*{-0.0cm}
\includegraphics[scale=0.62]{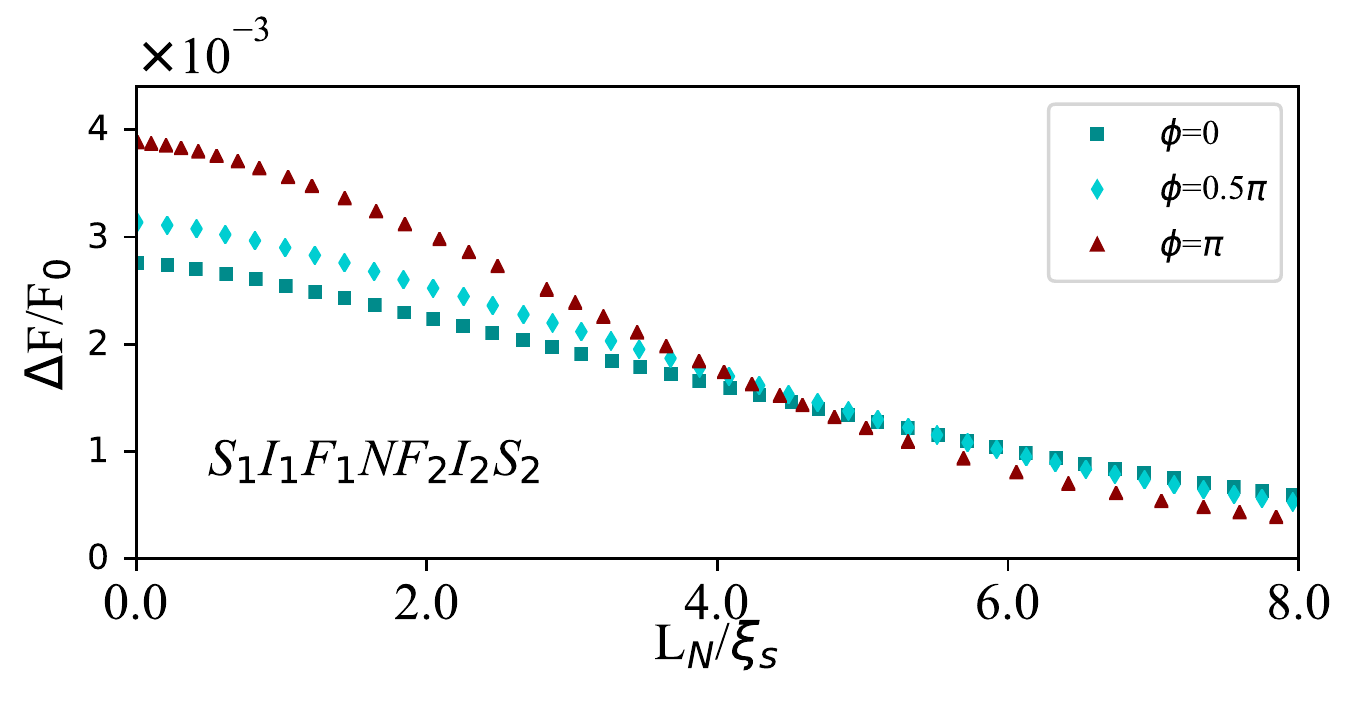}
\caption{\label{fig7} Exchange coupling of the junction, $\Delta$F=F$_{PP}$-F$_{AP}$, versus L$_N$/$\xi_s$ for the S$_1$/I$_1$/F$_1$/N/F$_2$/I$_2$/S$_2$ structure satisfying g$_{F_{1}S_{1}}$=0.1g$_{F_{2}S_{2}}$ with T/T$_c$=0.1, h/T$_c$=10, and L$_F$/$\xi_s$=0.1 manifesting that the asymmetric junction always favor anti-parallel configuration irrespective of the phase differences. Here, F$_0$ is the free energy of the system when the magnetizations are collinear.}
\end{figure}

In addition, the behavior of the exchange coupling for a Josephson system with asymmetric barriers is investigated. To be more precise, the S$_1$/I$_1$/F$_1$/N/F$_2$/I$_2$/S$_2$ structure with the contact with very different g$_{S_{1}F_{1}}$ and g$_{S_{2}F_{2}}$ conductances are taken into account. It is found that in contrast to what happens in symmetric S$_1$/F$_1$/N/F$_2$/S$_2$ systems, such nanojunction always favor anti-parallel configuration irrespective of the phase differences. As seen in Fig. \ref{fig7}, this finding is always true for all values of L$_N$/$\xi_s$.
\subsubsection{Josephson junctions with FS reservoirs}
\begin{figure}
\vspace*{-1.0cm}
\centering
\hspace*{-0.4cm}
\vspace*{-0.45cm}
\includegraphics[scale=0.62]{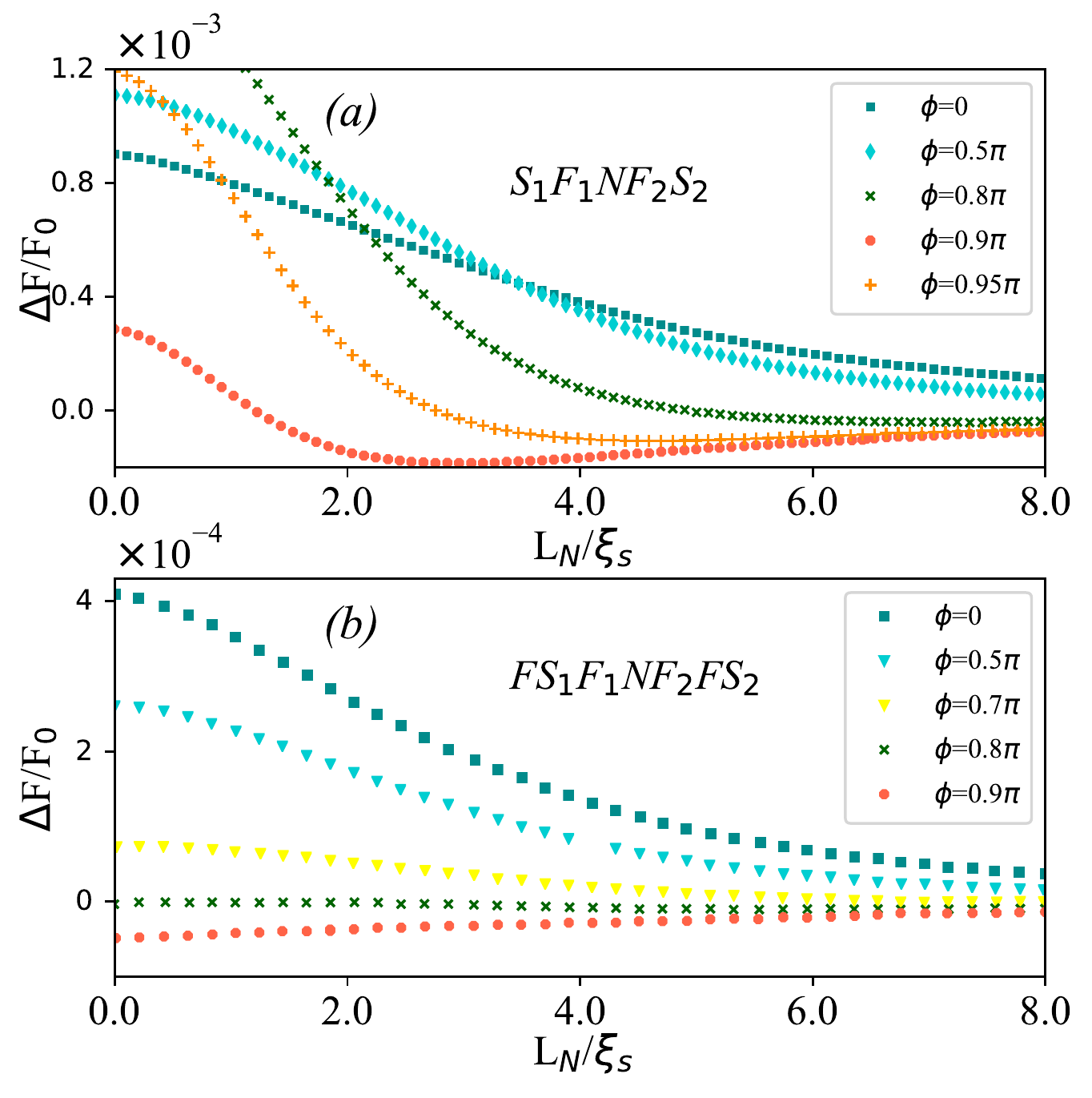}
\caption{\label{fig8}(a) Exchange coupling Difference of the junction, $\Delta$F=F$_{PP}$-F$_{AP}$, versus L$_N$/$\xi_s$ for S$_1$/F$_1$/N/F$_2$/S$_2$ nanosystem when L$_F$/$\xi_s$, h/T$_c$=10 and, T/T$_c$=0.1. (b) The same as (a) but for a FS$_1$/F$_1$/N/F$_2$/FS$_2$ structure presenting that the transition from antiparallel-parallel equilibrium configuration to the parallel-parallel one occurs in a very sharp slope for the FS$_1$/F$_1$/N/F$_2$/FS$_2$ system.}
\end{figure}

The other configuration in which the equilibrium exchange coupling is investigated is the Josephson junction consisting of two singlet ferromagnetic superconductor reservoirs in the form of FS$_1$/F$_1$/N/F$_2$/FS$_2$. The exchange field of the reservoirs makes an angle $\alpha$ with each other in similarity to the angle between the exchange fields of the F$_1$ and F$_2$. The Figs. \ref{fig8}(a) and \ref{fig8}(b) illustrate the behavior of the difference between the parallel and antiparallel configuration of the free energy. As seen in Fig. \ref{fig8}(a), at low temperatures, when T/T$_c$=0.1, h/T$_c$=0.1 and L/$\xi_F$=0.1, the FS$_1$/F$_1$/N/F$_2$/FS$_2$ structure manifests antiparallel-parallel equilibrium exchange coupling for the phase difference between $\phi$=0 and $\phi$=0.8$\pi$. This behavior holds whatever the length of the normal layer is. For larger superconducting phase differences, this manner reverses such that the equilibrium exchange coupling will be parallel-parallel configuration of the two FSs and the two ferromagnetic layers of the all values of the normal layer, L$_N$/$\xi_s$. In other words, the transition from antiparallel-parallel equilibrium configuration to the parallel-parallel one occurs in a very sharp slope. The achieved results are also compared with that of the nanosystem of S$_1$/F$_1$/N/F$_2$/S$_2$, where the structure starts to experience the $\alpha$=0 equilibrium for smaller values of $\phi$. Concretely, when $\phi$=0.8$\pi$ the free energy is minimum partially at $\alpha$=0 for L$_N$/$\xi_s$$\ge$5. As Fig. \ref{fig8}(b) shows, in contrast to the behavior of the $\Delta$F=F$_{PP}$-F$_{PAP}$ plot for FS$_1$/F$_1$/N/F$_2$/FS$_2$, it takes $\Delta\phi$=0.4$\pi$ starting from $\phi$=0.75$\pi$ for this plot to be completely negative for the whole values of the normal layer length, L$_N$/$\xi_s$.
\subsection{Ballistic triplet FS$_1$/N/FS$_2$ Josephson junction}
In this section the effect of the superconducting correlations on the STT behavior in ferromagnetic superconductor Josephson junctions of the form FS$_{1}$/N/FS$_{2}$ with FSs being chiral triplet p-wave ferromagnetic superconductors in similarity to the A-phase in liquid $^3$He with the pairing potential of the form $\Delta$=$\Delta_{0}$(k$_x$+i$k_y$)/$k_F$ are investigated. The chiral p-wave state is an analogue of the A-phase (Anderson-Brinkman- Morel phase) of the superfluid $^3$He in which intrinsically the time reversal symmetry is broken\cite{Volovik2000,Kallin2016,Etter2017}. Further, it is not claimed that the A-phase in liquid $^3$He is ferromagnetic. We have only established that such systems indicate p-wave order of parameters assumed as chiral p-wave superconducting gaps similar to the A-phase phase in liquid $^3$He. The existence of the in-plane exchange torque in addition to the equilibrium one mediated by Andreev states is reported. The magnetic torques are also found to be dependent to the spin-resolved phase differences of the superconducting order parameters as well as to an externally applied phase difference. Further, the formation of the spin-triplet electron-hole ABSs with a non-collinear spin polarization has an essential effect on the existed torques.
\subsubsection{Bound State Energies; E$<$0}
This section deals with the momentous study of the bound states. For subgap energies E$\le|\Delta^{\uparrow,\downarrow}|$, successive AR at the two N/FS interfaces and the coherent propagation of the excitations between the reflections lead to the establishment of the ABSs. In our FS$_1$/N/FS$_2$ system, ABSs are correlated electron-hole pairs in spin-triplet states with a non-collinear polarization. The spectrum of the ABSs is found by considering the electronic states in the two FS regions and N layer while matching the corresponding spinors and their derivatives at the right and the left N/FS-interfaces. The calculation is restricted to the case of a short NM contact with thickness L$_N$ that is much smaller than the superconducting coherence length $\xi_s$. Specifically speaking the value of $\frac{L_N}{\lambda_F}$ is taken to be 0.2, hence $\frac{L_N}{\xi_s}$ will be 0.2$\frac{\pi^2}{\Delta/E_F}$. With $\Delta/E_F$ being 0.01, the $\frac{L_N}{\xi_s}$ becomes $\approx$0.002. In this limit, the subgap Andreev states with |E|$\le$$\Delta^{\uparrow,\downarrow}$ dominate the superconducting transport properties. For the short length structures, as previously reported in Ref. [\onlinecite{Shomali2012}], the phase dependence for the branches of the bound state energy is shifted by $\pm\alpha$ with respect to that of a conventional short S/N/S junction \cite{Kulic1977} determined by $\epsilon_{e,h,\uparrow,\downarrow}$=$\pm\Delta_0\cos{[(\varphi\pm\alpha)/2]}$ while considering transport normal to the interface and $\Delta$$\theta$=0. On the other hand for the nano structures with $\Delta$$\theta$=$\pi$, the dependence on $\alpha$ is rather weak, such that the dependence is almost the same as the conventional S/N/S case \cite{Shomali2012}.

\subsubsection{Verification}
At the first step, the accuracy of the developed numerical template has been verified by implying to the reduced systems. The F$_1$/N/F$_2$, S$_1$/N/S$_2$, singlet FS$_1$/N/FS$_2$. As expected, only the out-of-plane STT $\tau_x$ appears and no evidence for the occurrence of the in-plane spin transfer torque have been found. Also, the triplet FS$_1$/N/FS$_2$ for only $k_y$=0 is investigated. It is obtained that only the equilibrium out-of-plane spin transfer torque is detectable for such system. At the next subsection we report the results for the study of the full triplet FS$_1$/N/FS$_2$ with integrating over possible $k_y$s. Interestingly, non-zero in-plane spin transfer torque are calculated. The finding suggest that the in-plane spint transfer torques, $\tau_{y,z}$, are presented due to the manifestation of the $ik_y$ in the superconducting gap. All the numerical implementation are accessible via \cite{Zahracode3}.

\subsubsection{In-plane/Out-of-plane spin transfer torque $\tau_{x,y,z}$}
\begin{figure}
\hspace*{-0.72cm}
\vspace*{-0.1cm}
\includegraphics[width=1.1\columnwidth]{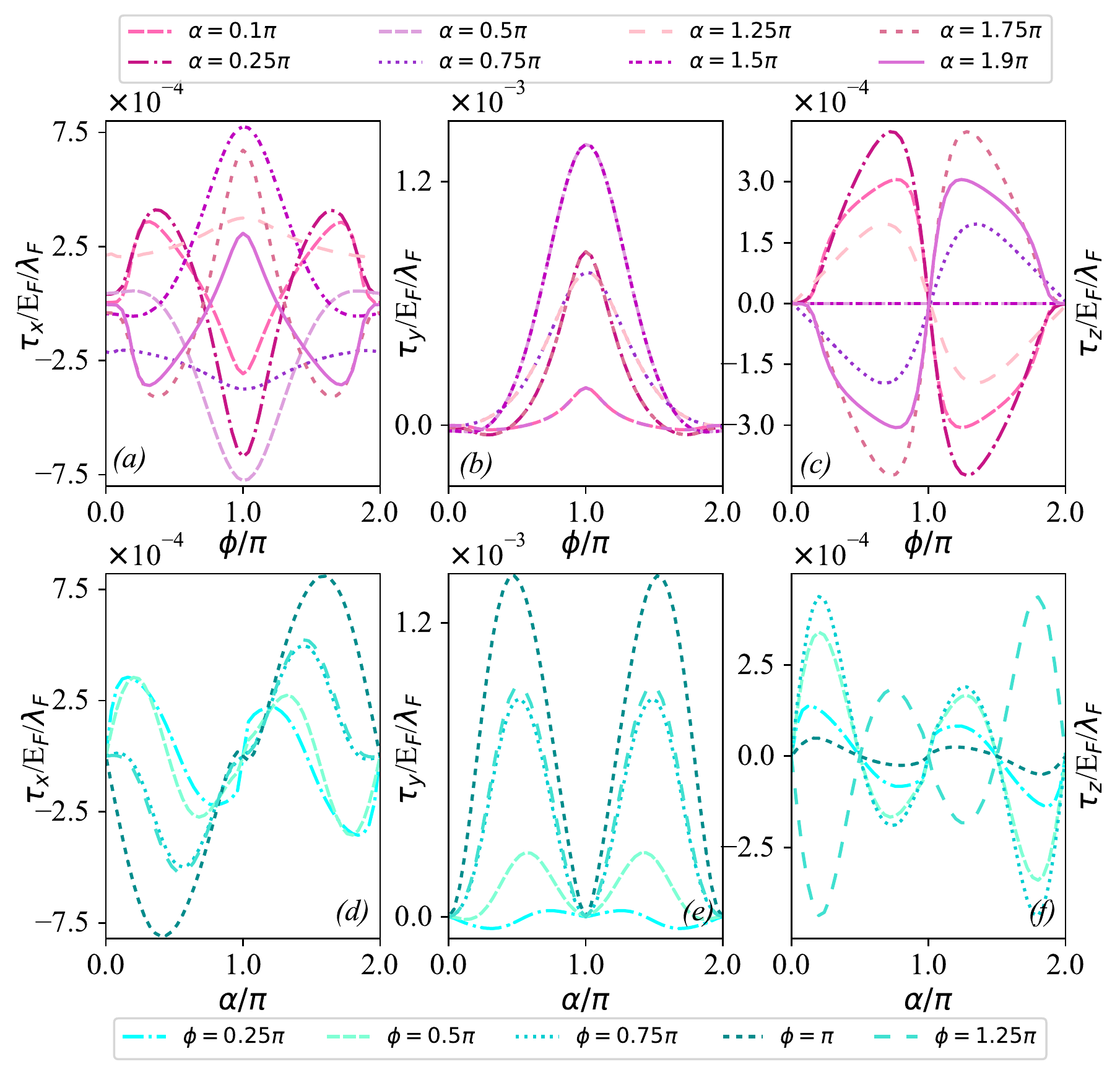}
\caption{\label{tauphi}(a), (b), and (c) are respectively $\tau_{x}$, $\tau_{y}$, and $\tau_{z}$ versus scaled superconducting phase difference, $\phi$/$\pi$, for different values of the angle between the exchange fields of the two FS reservoirs, $\alpha$ when $\Delta$$\theta$=0. STT in $x$-direction is even relative to $\phi$ and odd in $\alpha$ while $\tau_y$ and $\tau_z$ respectively are even and odd relative to both $\phi$ and $\alpha$. (d), (e), and (f) presenting $\tau_{x}$, $\tau_{y}$, and $\tau_{z}$ versus $\alpha$/$\pi$ for different values of the $\phi$ while $\Delta$$\theta$$_{1,2}$=0. All plots are drawn for x/$\lambda_F$=0.22. The results are for the conditions with h$_{{1,2}}$/E$_F$=0.1, $\Delta_{{1,2}}^{\uparrow(\downarrow)}$/E$_F$=0.01.}
\end{figure}

The spin-polarized Andreev states carrying charge and spin supercurrent lead to an equilibrium out-of-plane and the nonequilibrium in-plane exchange torque originating from superconducting correlations between the magnetization vectors of the two FSs. All three components of the STT will be calculated using Eq. (\ref{tausst}) and by summing over the contribution of all ABSs. The out-of-plane $\tau_x$ component which in the order of the magnitude is in good agreement with the recent theoretical work \cite{Halterman2015} tends to rotate the exchange coupling fields in $y$-$z$ directions around the $x$-axis, while the in-plane out-of-equilibrium tries to mislead the favorable configuration of the exchange coupling from $\alpha$=0 or $\alpha$=$\pi$. It is interesting to highlight that in contrast to the other heretofore studied Josephson junctions, the in-plane STT are also appeared. Interestingly, as seen in Fig. \ref{tauphi} when $\Delta$$\theta$=0,  it is obtained that the superconducting torque in $x$-direction is odd in $\alpha$ but even in $\varphi$, obeying the relations $\tau_x$($\alpha$,$\varphi$)=-$\tau_{x}$(2$\pi$-$\alpha$,$\varphi$). This is true while $\tau_y$ is even in $\alpha$ and $\phi$ fulfilling the relation $\tau_y$($\alpha$,$\varphi$)=$\tau_{y}$(2$\pi$-$\alpha$,$\varphi$). Further the $\tau$ in $z$-direction presents a behavior at odds with showing both odd manner relative to the $\alpha$ and the superconducting phase difference. Somehow the appeared in-plane torque can be the triplet Josephson supercurrent induced torque in analogy to the the nonequilibrium current-induced torque emerged for the normal metal FNF contact. 

\begin{figure}
\hspace*{-0.65cm}
\vspace*{-0.1cm}
\includegraphics[width=1.1\columnwidth]{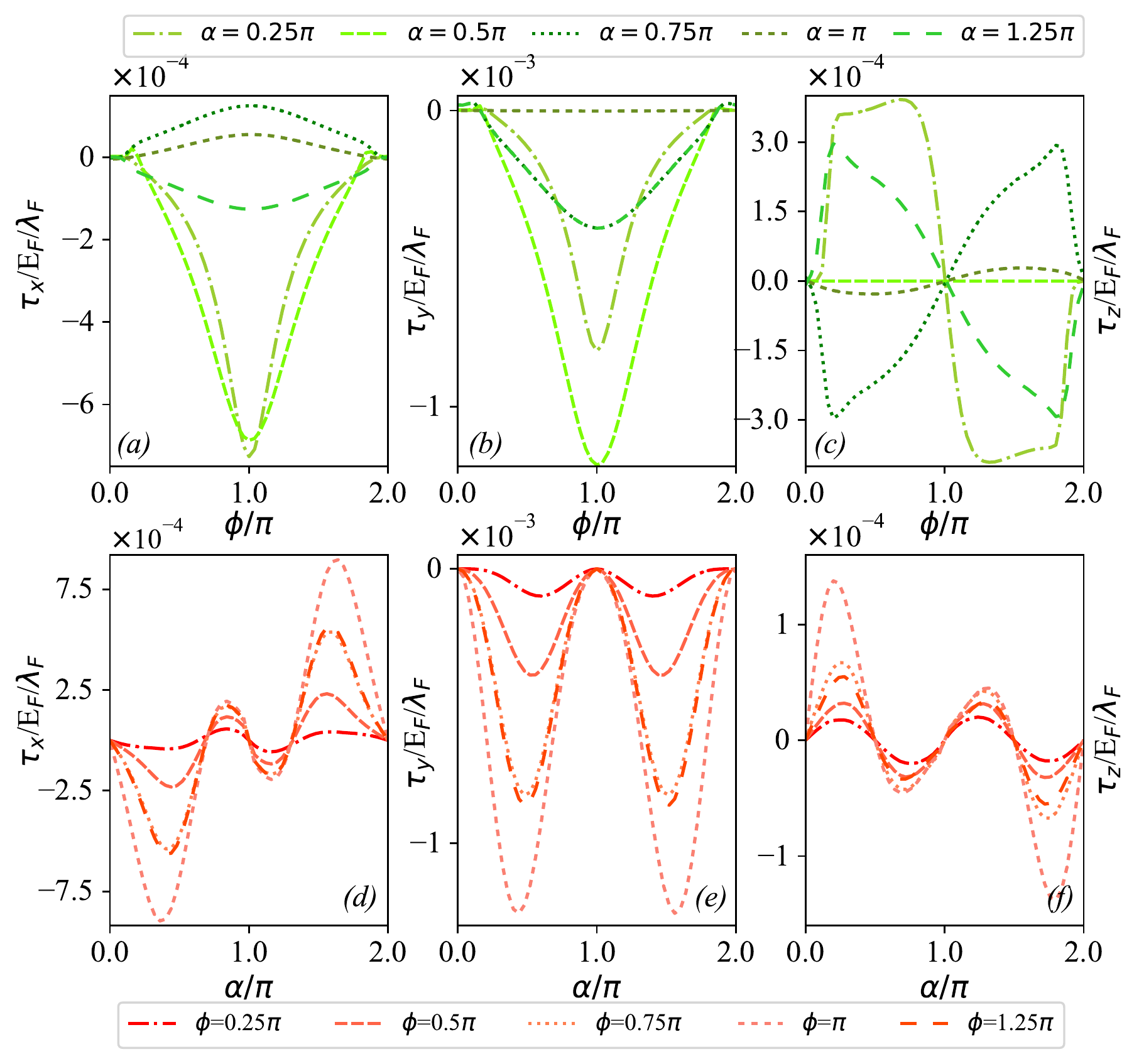}
\caption{\label{tauphi2}(a), (b), and (c) are respectively $\tau_{x}$, $\tau_{y}$, and $\tau_{z}$ versus scaled superconducting phase difference, $\phi$/$\pi$, for different values of the angle between the exchange fields of the two FS reservoirs, $\alpha$ when $\Delta$$\theta$$_{1,2}$=$\pi$. (d), (e), and (f) exhibiting $\tau_{x}$, $\tau_{y}$, and $\tau_{z}$ versus $\alpha$/$\pi$ for different values of $\phi$ while $\Delta$$\theta$$_{1,2}$=$\pi$. The symmetry behavior is the same as that of the case with $\Delta$$\theta$$_{1,2}$=0. The electrical properties of the system are the same as those implied in Fig. \ref{tauphi}.}
\end{figure}

On the other hand, the $\tau$-$\alpha$ and $\tau$-$\phi$ relations for the $\Delta$$\theta$=$\pi$ are also investigated. Fig. \ref{tauphi2} demonstrates such dependency when h$_{{1,2}}$/E$_F$=0.1, $\Delta_{{1,2}}^{\uparrow(\downarrow)}$/E$_F$=0.01 and L$_N$/$\lambda_F$=0.2. As it is obvious, the overall behavior of the STT in all three directions relative to the $\phi$ and $\alpha$ remains unchanged. Nevertheless, reversing the sign of the STT as $\Delta$ becomes equal to $0$ can be the field of interest. The importance of suchlike deportment lies in the previously mentioned requirement for the new generation transistor production. This sign changing satisfies the demand for the essential two different states revolving easily by only changing the phase between the majority and minority spin superconducting order parameters. On the other hand, the presence of the second harmonic in out-of-plane component of the torque while $\Delta$$\theta$)$_{1,2}$=0 in combination with the second harmonic appeared in $\tau_z$ for both $\Delta \theta$$_{1,2}$=0 and $\Delta$$\theta$$_{1,2}$=$\pi$ is also an intriguing obtained result.   
The featured outcome of the computation emerges while dealing with the demeanor of the STT inside the second ferromagnetic superconductor of the FS$_1$/N/FS$_2$.

\begin{figure}
\vspace*{-0.01cm}
\centering
\hspace*{-0.8cm}
\vspace*{-0.3cm}
\includegraphics[width=1.12\columnwidth]{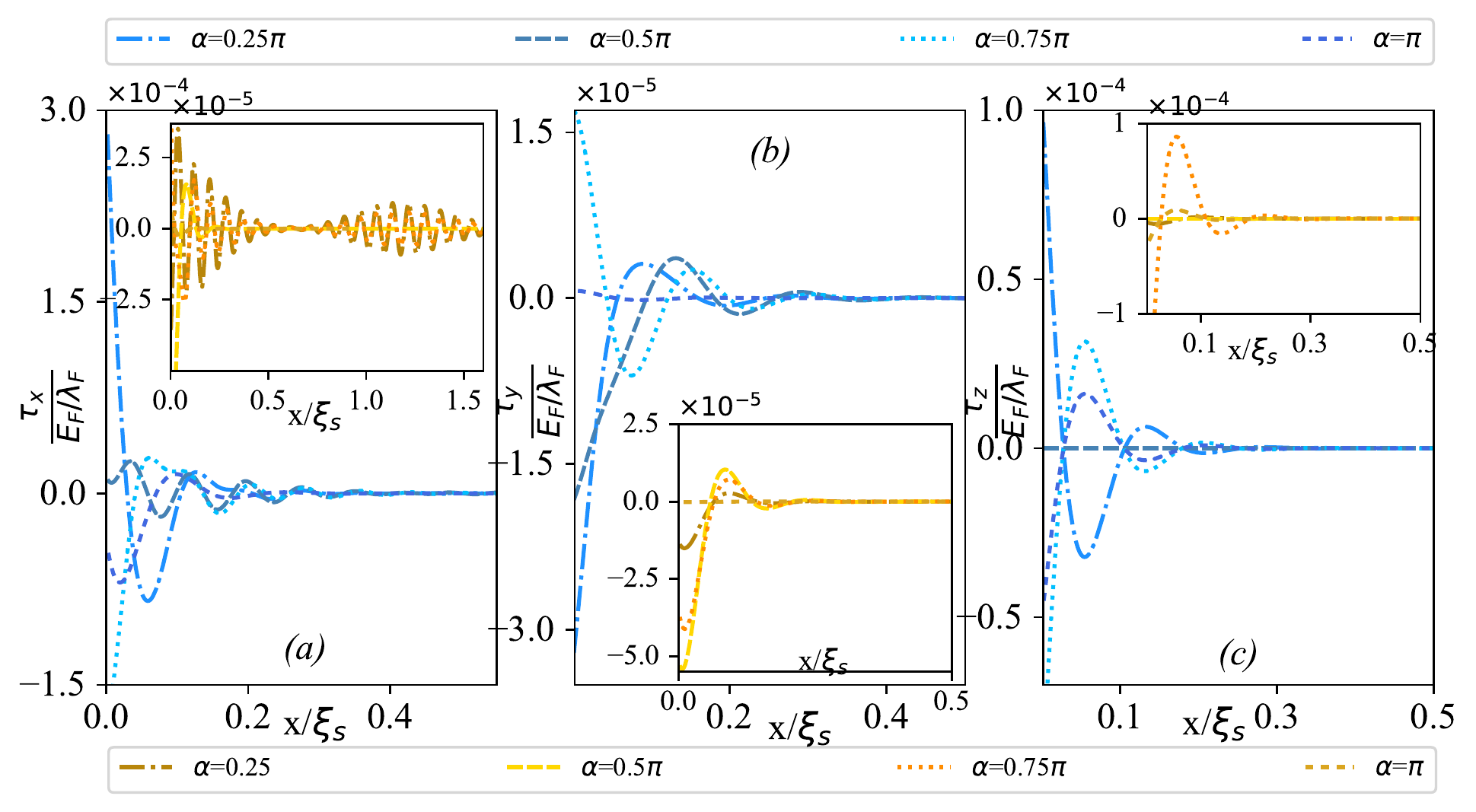}
\caption{\label{tauvln}(a) $\tau_{x}$ versus $x$/$\xi_s$, which is the coordinate of the position in the FS$_2$ when $\phi$/$\pi$=0.25 and $\Delta$$\theta$$_{1,2}$=0 for different values of $\alpha$/$\pi$. The inset presents such behavior for $\Delta$$\theta$=$\pi$. The existence of two different periods in out-of-plane STT when $\Delta$=$\pi$ is manifest in inset. The value of $\tau_x$ emerging with two oscillation periods drops notably for $\Delta$=$\pi$. (b) and (c) are the same as (a) but respectively for $\tau_{y}$ and $\tau_z$, mainfestig the value of the in-plane STT is nearly the same for both $\Delta$=$\pi$ and $\Delta$=0. The outcomes establishes available with h$_{{1,2}}$/E$_F$=0.1, $\Delta_{{1,2}}^{\uparrow(\downarrow)}$/E$_F$=0.01.}
\end{figure}

Fig. \ref{tauvln}, denotes the behavior of the STT in the second ferromagnetic superconductor (FS$_2$) for $x$$>$L$_N$. In both cases, $\tau_{x,y,z}$ shows spatial oscillations with a period that is roughly proportional to the 1/$h$. The amplitude of $\tau_{x,y,z}$ exposes a decay with $x$ over a penetration length which is a fraction of $\xi$. For both values of $\Delta$$ \theta$$_{1,2}$ the amplitude as well as the direction of the torque can be tuned by externally changing the phase difference over the contact.
Fig. \ref{tauvln} shows the behavior of the integrated over $k_y$ STT in the second ferromagnetic superconductor (FS$_2$), in the places where $x$$\ge$L$_N$. Figs. \ref{tauvln}(a), (b) and (c) respectively demonstrate the decayed oscillatory behavior of the out-of-plane $\tau_x$ and in-plane $\tau_{y,z}$ when $\Delta$$\theta$$_{1,2}$=0. There exists a good consistency between the results obtained for the out-of-plane spin transfer torque and the outcomes from the previous literatures \cite{Zhao2008,Alidoust2010,Halterman2015}. As seen in Fig. \ref{tauvln}, this fading vacillation comportment strongly depends on the angle and the superconducting phase difference between the two FSs. Further other parameters of the exchange coupling and superconducting gap are also crucial in delineating the STT in all three directions. Therewith, the insets of Figs. \ref{tauvln}(a), (b) and (c) compares the obtained results for $\Delta$$\theta$$_{1,2}$=$\pi$ with that of the $\Delta$$\theta$$_{1,2}$=0. At first glance, it is eminent that the amplitude of the exchange torque when $\Delta$$\theta$$_{1,2}$ is equal to $\pi$ decreases notably relative to the situation with $\Delta$$\theta$$_{1,2}$=0. The other absorbing observation is the apparition of the two different oscillation periods. As the inset of Fig. \ref{tauvln}(a) suggests, this finding ascertains that the exchange torque is the combination of the two torques with two different far from periods of oscillation which unify reaching the angle $\alpha$/$\pi$=0.5 and getting away again while increasing $\alpha$ more. 

\begin{figure}
\vspace*{-0.21cm}
\hspace*{-0.68cm}
\vspace*{-0.33cm}
\includegraphics[width=1.1\columnwidth]{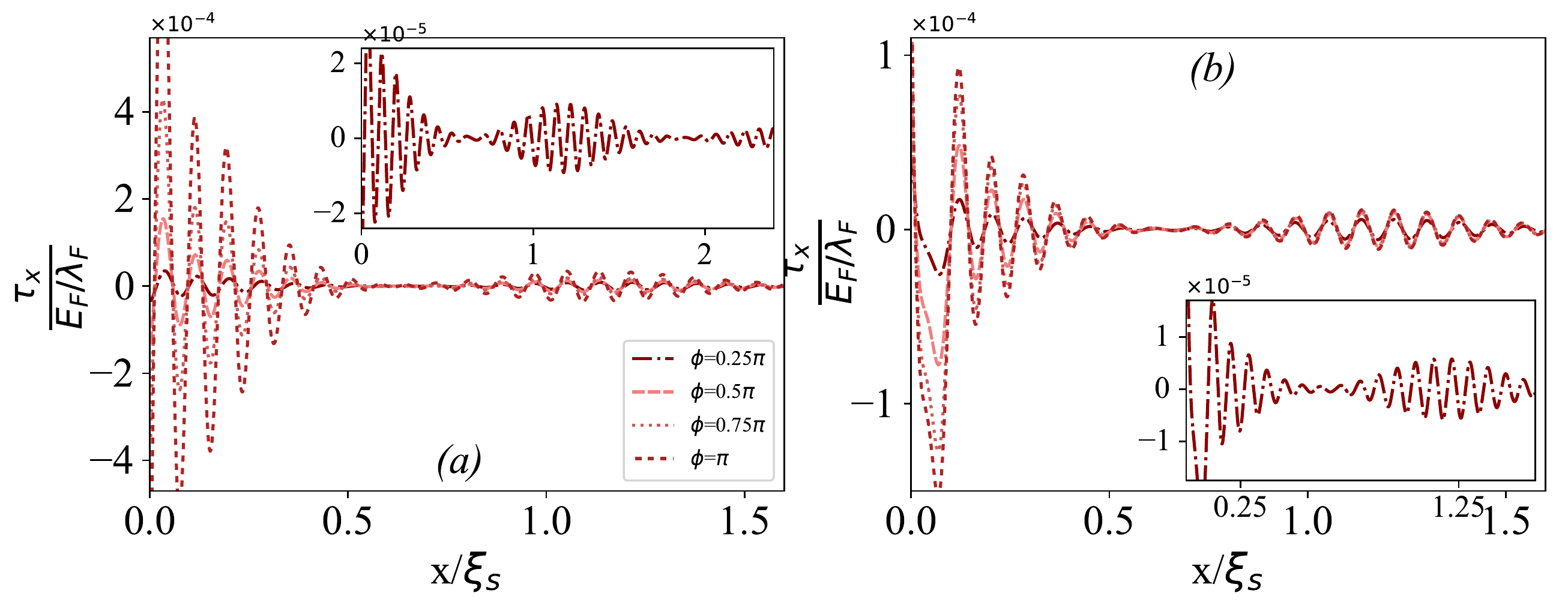}
\caption{\label{alpha} (a) Out-of-plane STT versus length for different values of $\phi$/$\pi$ when $\alpha$=0.25$\pi$. The beat like STT behavior appears when $\alpha$ is an odd multiple of $\pi$/4. The inset is the more visible zoomed plot of the more conspicuous beat like behavior happening for $\phi$=0.25$\pi$. (b) and the inset in, is the same as (a) but for $\alpha$=0.75$\pi$ while h$_{{1,2}}$/E$_F$=0.1 and $\Delta_{{1,2}}^{\uparrow(\downarrow)}$/E$_F$=0.01.}
\end{figure}

The more detailed behavior is shown in Fig. \ref{alpha}. As Figs. \ref{alpha}(a) and (b) suggest for the $\Delta$$\theta$=$\pi$, the overall behavior of the out-of-plane STT presenting beep like trend depends on the angle between the exchange fields of the FSs and also interestingly on the superconducting phase difference. It is shown that the two characteristic oscillation periods occur when $\alpha$ is an odd multiple of the $\pi$/4. Also, this behavior is more likely to happen when $\phi$ is equal to the 0.25$\pi$. The insets of Figs. \ref{alpha}(a) and (b) are the zoomed presentation of the out-of-plane STT for $\phi$/$\pi$=0.25, respectively, when $\alpha$/$\pi$ is 0.25 and 0.75.
\section{Conclusion}
\label{sec:4}
The ferromagnetic Josephson junctions including the nanosystems with the triplet/singlet ferromagnetic superconductor reservoirs have been investigated. First, we have studied the diffusive magnetic junctions of S$_1$/F$_{1}$/I$_1$/N/I$_2$/F$_2$/S$_2$ structure with different insulator barriers (I). The spin transfer torque, in similarity to the other diffusive Josephson junctions, only appears in the normal direction to the plane of the exchange fields of the F$_1$ and F$_2$. The out-of-plane spin transfer torque (STT) results in the parallel or antiparallel configuration of the minimum of the free energy. Also, the antiparallel/parallel or vice versa parallel/antiparallel transition of the favorable configuration of the exchange coupling occurs which depends on the parameters of the nanostructure such as the exchange field, h/T$_c$ and the temperature relative to the critical temperature, T/T$_c$. The relative width defined as the ratio of $\Delta$$\phi$, the interval in which the transition is happening, relative to $\phi_i$, being the phase difference in which the transition starts, of the S$_1$/F$_1$/N/F$_2$/S$_2$, symmetric S$_1$/F$_1$/I/N/I/F$_2$/S$_2$ and symmetric/asymmetric S$_1$/I/F$_1$/N/F$_2$/I/S$_2$ double barrier junctions were investigated. In practice, the temperature and the exchange field are fixed and the value of $\phi$ is varied for detection of possible change in the favorable configuration of the Josephson junctions. In more details the free energy of the nano structures has plotted versus the angle between the exchange couplings $\alpha$. We have obtained that the minimum of this free energy as a function of the $\alpha$ always occurs when $\alpha$=0 or $\alpha$=$\pi$ depending on the other involved parameters. Furthermore, for asymmetric barriers as the exchange field increases or as the temperature becomes higher, the interval of phase differences, in which the absolute minimum of the free energy as a function of the $\alpha$ occurs in $\alpha$=$\pi$ gets larger. When the exchange field reaches the specified value, the interval starts to become smaller until it gets near to the saturation point. Moreover, the Ballistic triplet FS$_{1}$/N/FS$_{2}$ is studied by solving the BDG Equation. To be more clear, we have studied from the primitive structure to the more complex one. In more details, first the F$_1$/N/F$_2$, S$_1$/N/S$_2$, singlet FS$_1$/N/FS$_2$, the triplet FS$_1$/N/FS$_2$ with $k_y$=0 are studied. As it is seen in Table. \ref{Tab2-BP}, for these system the out-of-plane STT $\tau_x$ occurs and no evidence for the occurrence of the in-plane spin transfer torque have been found. At the second stage the full triplet FS$_1$/N/FS$_2$ with integrating over possible $k_y$s is investigated. Interestingly for the triplet FS$_1$/N/FS$_2$ with $k_y\neq$0, we have obtained that in addition to the existence of the out-of-plane STT, the in-plane STT is appeared along the direction of the exchange fields. The evidences present that the appearance of the non-zero k$_y$ in the form of the paring potential is responsible for the occurrence of such novel spin transfer torque. The apparition of such behavior suggests that the minimum of the free energy may occur while the exchange fields of the ferromagnets make the relative angle different from $\alpha$=0 or $\alpha$=$\pi$. We believe that the present study is the first work announcing the occurrence of the in-plane STT. The appearance of the STT with two oscillation periods inside the ferromagnetic superconductor depending on the difference between the majority and minority spin superconducting order parameters, is the other compelling uncovering of this study.\\

\begin{table*}[htbp]
\caption{The spin transfer torque in different types of ballistic Josephson junctions \newline}

 \label{Tab2-BP}
\centering
\begin{small}
\hspace*{-0.4cm}
\begin{tabular}{|c|c|c|c|c|}
  \hline
Studied structure &  Superconducting gap; $\Delta$ & Out-of-plane $\tau_x$ & 
 In-plane $\tau_{y,z}$   \\ \hline
F$_1$/N/F$_2$   & 0 & Yes & No  \\ \hline
S$_1$/N/S$_2$ & $|\Delta_0| e^{i\phi}$ & No & No \\ \hline
FS$_1$/N/FS$_2$ & $|\Delta_0| e^{i\phi}$ & Yes & No  \\ \hline
FS$_1$/N/FS$_2$; k$y$=0 & $|\Delta_0| e^{\pm i\theta^{\uparrow\downarrow}} e^{i\phi}$ & Yes & No  \\ \hline
FS$_1$/N/FS$_2$; k$y$$\neq$0  & $ |\Delta_0| \frac{\pm k_x^+ \pm i k_y}{\sqrt{k^+=(k_x^+)^2+(k_y)^2}}e^{\pm i\theta^{\uparrow\downarrow\sigma}}$e$^{i\phi}$, & Yes & Yes  \\ \hline
\end{tabular}
\end{small}
\end{table*}

\appendix
\section{Bogoliubov-de-Gennes formalism for F$S_{1}$/N/F$S_{2}$}
For the ferromagnetic superconductor (FS) with the magnetization along the $y$-$z$ axis it is obtained that,
\begin{widetext}
\hspace*{-0.5cm}
$\left(
\begin{array}{cccc}
-\frac{\nabla^2}{2m}-E_F^{(\ell)}-h_\ell & 0 & \Delta_{\ell e(h)}^{\uparrow} & 0 \\
0 & -\frac{\nabla^2}{2m}-E_F^{(\ell)}+h_\ell & 0 & \Delta_{\ell e(h)}^{\downarrow} \\
\Delta_{\ell e(h)}^{\uparrow *} & 0 & \frac{\nabla^2}{2m}+E_F^{(\ell)}+h_\ell & 0 \\
0 & \Delta_{\ell e(h)}^{\downarrow *} & 0 & \frac{\nabla^2}{2m}+E_F^{(\ell)}-h_\ell \\
\end{array}
\right)\left(
\begin{array}{c}
U_{\ell e(h)}^{\uparrow} \\
U_{\ell e(h)}^{\downarrow} \\
V_{\ell e(h)}^{\uparrow} \\
V_{\ell e(h)}^{\downarrow} \\
\end{array}
\right)=E\left(\begin{array}{c}
U_{\ell e(h)}^{\uparrow} \\
U_{\ell e(h)}^{\downarrow} \\
V_{\ell e(h)}^{\uparrow} \\
V_{\ell e(h)}^{\downarrow} \\
\end{array}
\right).$
\end{widetext}
Here, $\ell$={1,2} denotes the parameters for the first and the second ferromagnetic superconductor respectively. Electrons and holes are presented as e,h. Also, $\sigma$=+1,-1 indicates the spin up and spin down electron and holes. All the solutions of the above equation are written as U$_{\ell e(h)}^{\sigma}$=u$_{\ell e(h)}^\sigma$ e$^{i q_{\ell e(h)}^\sigma x}$,V$_{\ell e(h)}^{\sigma}$=v$_{\ell e(h)}^\sigma$ e$^{i q_{le(h)}^\sigma x}$ and consequently the above equation takes the following form,
\begin{multline}
\hspace*{-0.5cm}
\left(
\begin{array}{cc}
-\frac{\nabla^2}{2m}-E_F^{(\ell)}-{\bf{\sigma}}\cdot{\bf{h}_\ell}-E & \Delta_{\ell e(h)}^{\sigma} \\
\Delta_{\ell e(h)}^{\sigma*} & \frac{\nabla^2}{2m}+E_F^{(\ell)}+{\bf{\sigma}}\cdot{\bf{h}_\ell}-E \\
\end{array}
\right) \\ \left(
\begin{array}{c}
u_{\ell e(h)}^\sigma \\
v_{\ell e(h)}^\sigma \\
\end{array}
\right)e^{i q_{\ell e(h)}^\sigma x}e^{i k_y y}=0.
\end{multline}
Applying Hamiltonian on the wave function it is attained that,
\begin{widetext}
\begin{equation}
\label{A4}
\hspace*{-0.5cm}
\left(
\begin{array}{cc}
\frac{(q_{\ell e(h)}^\sigma)^2}{2m}+\frac{k_y^2}{2m}-E_F^{\ell}-{\bf{\sigma}}\cdot{\bf{h}_\ell}-E & \Delta_{\ell e(h)}^{\sigma} \\
\Delta_{\ell e(h)}^{\sigma*} & -\frac{(q_{\ell e(h)}^\sigma)^2}{2m}-\frac{k_y^2}{2m}+E_F^\ell+{\bf{\sigma}}\cdot{\bf{h}_\ell}-E \\
\end{array}
\right)
\left(
\begin{array}{cc}
u_{\ell e(h)}\sigma \\
v_{\ell e(h)}\sigma \\
\end{array}
\right)=0;
\end{equation}
\begin{equation}
 A=\left(
\begin{array}{cc}
\frac{(q_{\ell e(h)}^\sigma)^2}{2m}+\frac{k_y^2}{2m}-E_F^{\ell}-{\bf{\sigma}}\cdot{\bf{h}_\ell}-E & \Delta_{\ell e(h)}^{\sigma} \\
\Delta_{\ell e(h)}^{\sigma*} & -\frac{(q_{\ell e(h)}^\sigma)^2}{2m}-\frac{k_y^2}{2m}+E_F^\ell+{\bf{\sigma}}\cdot{\bf{h}_\ell}-E \\
\end{array}
\right).
\end{equation}
\end{widetext}
The values of q$_{\ell e(h)}^\sigma$ as the functions of E are found by letting the determinant of A equal to zero,
\begin{multline}
\nonumber \hspace*{-0.3cm} det A=(\frac{(q_{\ell e(h)}^\sigma)^2}{2m}+\frac{k_y^2}{2m}-E_F^{(\ell)}-{\bf{\sigma}}\cdot{\bf{h}_\ell}-E)\\-(\frac{(q_{\ell e(h)}^\sigma)^2}{2m}+\frac{k_y^2}{2m}-E_F^{(\ell)}-\sigma.h_{\ell}+E)+|\Delta_{\ell e(h)}^{\sigma}|^2=0.
\end{multline}
\begin{equation}
\nonumber
\hspace*{-0.3cm} \Rightarrow -\frac{(q_{\ell e(h)}^\sigma)^2}{2m}-\frac{k_y^2}{2m}+E_F^{(\ell)}+{\bf{\sigma}}\cdot{\bf{h}_\ell}=\pm \sqrt {E^2-|\Delta_{\ell e(h)}^{\sigma}|^2}
\end{equation}
\begin{equation}
\label{qleh}
\hspace*{-0.3cm} q_{\ell e(h)}^\sigma=\sqrt{2m(E_F^{(\ell)}+\sigma.h_{\ell}\pm \sqrt {E^2-|\Delta_{\ell e(h)}^{\sigma}|^2})-k_y^2}
\end{equation}
The $\pm$ sign in the above equation manifests two solutions related to the hole-like, q$_{\ell h}^\sigma$, and electron-like, q$_{\ell e}^\sigma$, solutions of $q$. At the next step, u$_{\ell e(h)}^{\sigma}$,v$_{\ell e(h)}^{\sigma}$ parameters will be attained. For each ferromagnetic superconductor in both spin direction, there are two sets of u,v parameters called as electron-like, u$_{\ell e}^{\sigma}$, and hole-like, u$_{\ell h}^{\sigma}$, solutions. For these specified estates by applying Eq. (\ref{qleh}) in matrix Eq. \ref{A4}, two following equations are obtained
\begin{equation}
\hspace*{-0.8cm}\Delta_{\ell e(h)}^{\sigma}v_{\ell e(h)}^\sigma-(E\mp\sqrt {E^2-|\Delta_{\ell e(h)}^{\sigma}|^2}) u_{\ell e(h)}^\sigma=0,
\end{equation}
and
\begin{equation}
\label{delta}
\hspace*{-0.8cm}\Delta_{\ell e(h)}^{\sigma*} u_{\ell e(h)}^\sigma-(E\pm\sqrt {E^2-|\Delta_{\ell e(h)}^{\sigma}|^2}) v_{\ell e(h)}^\sigma=0.
\end{equation}
Also the normalization condition will be considered,
\begin{equation}
\label{normalization}
\hspace*{-0.8cm} |u_{\ell e(h)}^\sigma|^2+|v_{\ell e(h)}^\sigma|^2=1.
\end{equation}
Now with consideration of the electron-like solutions of Eq. (\ref{delta}) and Eq. (\ref{normalization}) by applying the electron-like Eq. \ref{delta}, it is seen that
\begin{equation}
\label{eq:11}
\hspace*{-0.8cm} v_{\ell e}^\sigma=\frac{\Delta_{\ell e}^{\sigma*} u_{\ell e}^\sigma}{(E+\sqrt {E^2-|\Delta_{\ell e}^{\sigma}|^2})}.
\end{equation}
Substituting the obtained value of $v_{\ell e}^{\sigma}$ in the Eq. (\ref{normalization}) gives
\begin{equation}
\nonumber \hspace*{-0.8cm} |u_{\ell e}^\sigma|^2(1+\frac{|\Delta_{\ell e}^{\sigma}|^2}{(E+\sqrt {E^2-|\Delta_{\ell e}^{\sigma}|^2})^2})=1,
\end{equation}
\begin{equation*}
\nonumber \hspace*{-0.2cm} \Rightarrow |u_{\ell e}^\sigma|^2(\frac{E^2+E^2-|\Delta_{\ell e}^{\sigma}|^2+2E\sqrt {E^2-|\Delta_{\ell e}^{\sigma}|^2}+|\Delta_{\ell e}^{\sigma}|^2}{(E+\sqrt {E^2-|\Delta_{\ell e}^{\sigma}|^2})^2})=1,
\end{equation*}
\begin{multline}
\nonumber \hspace*{-0.4cm} \Rightarrow |u_{\ell e}^\sigma|^2(\frac{2E^2+2E\sqrt {E^2-|\Delta_{\ell e}^{\sigma}|^2}}{(E+\sqrt {E^2-|\Delta_{\ell e}^{\sigma}|^2})^2})= \\ |u_{\ell e}^\sigma|^2[\frac{2E(E+\sqrt {E^2-|\Delta_{\ell e}^{\sigma}|^2})}{(E+\sqrt {E^2-|\Delta_{\ell e}^{\sigma}|^2})^2}]=1,
\end{multline}
\begin{equation}
\nonumber \hspace*{-0.8cm} |u_{\ell e}^\sigma|^2[\frac{2E}{(E+\sqrt {E^2-|\Delta_{\ell e}^{\sigma}|^2})}]=1,
\end{equation}
\begin{equation}
\nonumber \hspace*{-0.2cm} |u_{\ell e}^\sigma|^2=[\frac{(E+\sqrt {E^2-|\Delta_{\ell e}^{\sigma}|^2})}{2E}]=[1+\frac{\sqrt{E^2-|\Delta_{\ell e}^{\sigma}|^2}}{E}]/2,
\end{equation}
\begin{equation}
\label{eq:12}
\hspace*{-0.8cm}\Rightarrow u_{\ell e}^\sigma=\sqrt{(1+\frac{\sqrt {E^2-|\Delta_{\ell e}^{\sigma}|^2}}{E})/2}.
\end{equation}
Using Eq. (\ref{eq:12}) and Eq. (\ref{eq:11}) the following result is acquired,
\begin{equation}
\label{eq.v}
\hspace*{-0.6cm} \Rightarrow v_{\ell e}^{\sigma}=\frac{\Delta_{\ell e}^{\sigma*}}{|\Delta_{\ell e}^{\sigma}|}\sqrt{(1-\frac{\sqrt {E^2-|\Delta_{\ell e}^{\sigma}|^2}}{E})/2}.
\end{equation}
Now the hole-like forms of Eq. (\ref{delta}) and Eq. (\ref{normalization}) are figured out to find the hole-like solutions,
\begin{equation}
\hspace*{-0.8cm} \Delta_{\ell h}^{\sigma*} u_{\ell h}^\sigma-(E-\sqrt {E^2-|\Delta_{\ell h}^{\sigma}|^2}) v_{\ell h}^\sigma=0
\end{equation}
\begin{equation}
\hspace*{-0.8cm} \nonumber v_{\ell h}^\sigma=\Delta_{\ell h}^{\sigma*} \frac{u_{\ell h}^\sigma}{(E-\sqrt {E^2-|\Delta_{\ell h}^{\sigma}|^2})}.
\end{equation}
Implying the normalization condition the following relation is acquired,
\begin{equation}
\nonumber \hspace*{-0.8cm} |u_{\ell e}^\sigma|^2(1+\frac{|\Delta_{\ell e}^{\sigma}|^2}{(E-\sqrt {E^2-|\Delta_{\ell e}^{\sigma}|^2})^2})=1.
\end{equation}
The performed calculation for the electron-like solutions should be repeated to accede to the pursuant formula,
\begin{equation}
\hspace*{-0.8cm} \nonumber u_{\ell h}^\sigma=\sqrt{(1-\frac{\sqrt {E^2-|\Delta_{\ell h}^{\sigma}|^2}}{E})/2}.
\end{equation}
\begin{equation}
\hspace*{-0.8cm} \Rightarrow v_{\ell h}^\sigma=\frac{|\Delta_{\ell h}^{\sigma*}|}{\Delta_{\ell h}^{\sigma}}\sqrt{(1+\frac{\sqrt {E^2-|\Delta_{\ell h}^{\sigma}|^2}}{E})/2}.
\end{equation}
$\frac{|\Delta_{\ell h}^{\sigma*}|}{\Delta_{\ell h}^{\sigma}}$ denotes the phase of the gap, so inevitably the above solutions can be written in the subsequent way,
\begin{equation}
\hspace*{-0.8cm}u_{\ell h}^\sigma=\frac{|\Delta_{\ell h}^{\sigma}|}{\Delta_{\ell h}^{\sigma*}}\sqrt{(1-\frac{\sqrt {E^2-|\Delta_{\ell h}^{\sigma}|^2}}{E})/2}
\end{equation}
\begin{equation}
\hspace*{-0.8cm}v_{\ell h}^\sigma=\sqrt{(1+\frac{\sqrt {E^2-|\Delta_{\ell h}^{\sigma}|^2}}{E})/2}=u_{\ell h}^\sigma.
\end{equation}
In the following, we come back to the studied case of the Josephson system consisting of the normal contact between two triplet FSs, schematically presented in Fig. \ref{geometry2}. The exchange field of FS$_1$ makes an angle $\alpha$ with that of the FS$_2$ while the phase difference between the two mentioned FSs is $\phi$. The chiral p-wave superconducting gaps demonstrated as $\Delta_{\ell e(h)}^{\sigma}$=$\Delta_{\ell \sigma,0}$(k$_{x(e,h)}$+i$k_y$)/$k_F^N$ are considered. All three FS$_1$, N, and FS$_2$ are taken to have the same fermi energy, E$_F^{(1)}$=E$_F^{N}$=E$_F^{(2)}$=E$_F$. The total wave function in the non-superconducting region is written as,
\begin{equation}
\hspace*{-0.8cm}
\Psi_{N}=
\left (
\begin{array}{rl}
\nonumber c_{1+}e^{ik^+x}+c_{1-}e^{-ik^+x}\\
\nonumber c_{2+}e^{ik^+x}+c_{2-}e^{-ik^+x}\\
\nonumber c_{3-}e^{ik^-x}+c_{3+}e^{-ik^-x}\\
c_{4-}e^{ik^-x}+c_{4+}e^{-ik^-x}\\
\end{array}
\right ).
\label{1}
\end{equation}
While for the first ferromagnetic superconductor FS$_1$, we have
\begin{multline}
\hspace*{-0.5cm} \Psi_{FS_1}=\hat{T}(
\left(
\begin{array}{c}
u_{1e}^{\uparrow} \\
0 \\
v_{1e}^{\uparrow}e^{-i \gamma^\uparrow_{1+}} \\
0 \\
\end{array}
\right)e^{iq_{1e}^\uparrow x}+r_e^\uparrow \left(
\begin{array}{c}
u_{1e}^{\uparrow} \\
0 \\
v_{1e}^{\uparrow}e^{-i \gamma^\uparrow_{1-}} \\
0 \\
\end{array}
\right)e^{-iq_{1e}^\uparrow x}\\+r_e^\downarrow\left(
\begin{array}{c}
0 \\
u_{1e}^{\downarrow} \\
0 \\
v_{1e}^{\downarrow}e^{-i \gamma^\downarrow_{1-}} \\
\end{array}
\right)e^{-iq_{1e}^\downarrow x}\\+r_h^\uparrow\left(
\begin{array}{c}
v_{1e}^{\uparrow}e^{i{\bar \gamma}^\uparrow_{1+}} \\
0 \\
u_{1e}^{\uparrow} \\
0 \\
\end{array}
\right)e^{iq_{1h}^\uparrow x}+r_h^\downarrow\left(
\begin{array}{c}
0 \\
v_{1e}^{\downarrow}e^{i{\bar \gamma}^\downarrow_{1+}} \\
0 \\
u_{1e}^{\downarrow} \\
\end{array}
\right)e^{iq_{1h}^\downarrow x})
\label{2n}
\end{multline}

\begin{equation}
\hspace*{-0.4cm}\hat{T}=\left(
\begin{array}{cccc}
\cos{\alpha/2} & i \sin{\alpha/2} & 0 & 0 \\
i \sin{\alpha/2} & \cos{\alpha/2} & 0 & 0 \\
0 & 0 & \cos{\alpha/2} & -i \sin{\alpha/2} \\
0 & 0 & -i \sin{\alpha/2} & \cos{\alpha/2} \\
\end{array}
\right).
\end{equation}
And the wave function for the FS$_2$ is,
\begin{multline}
\label{3}
\hspace*{-0.4cm}\Psi_{FS_2}= t_e^\uparrow \left(
\begin{array}{c}
u_{2e}^{\uparrow} \\
0 \\
v_{2e}^{\uparrow}e^{-i \gamma^\uparrow_{2+}} \\
0 \\
\end{array}
\right)e^{iq_{2e}^\uparrow x}+t_e^\downarrow\left(
\begin{array}{c}
0 \\
u_{2e}^{\downarrow} \\
0 \\
v_{2e}^{\downarrow}e^{-i \gamma^\downarrow_{2+}} \\
\end{array}
\right)e^{iq_{2e}^\downarrow x} \\ +t_h^\uparrow\left(
\begin{array}{c}
v_{2e}^{\uparrow}e^{i{\bar \gamma}^\uparrow_{2-}} \\
0 \\
u_{2e}^{\uparrow} \\
0 \\
\end{array}
\right)e^{-iq_{2h}^\uparrow x}+t_h^\downarrow\left(
\begin{array}{c}
0 \\
v_{2e}^{\downarrow}e^{i{\bar \gamma}^\downarrow_{2-}} \\
0 \\
u_{2e}^{\downarrow} \\
\end{array}
\right)e^{-iq_{2h}^\downarrow x}
\end{multline}
with r$_{e(h)}^{\uparrow(\downarrow)}$ and t$_{e(h)}^{\uparrow(\downarrow)}$ being respectively the reflection and the transmission coefficients. Also the following relation holds,
\begin{equation}
\hspace*{-0.8cm}k_x^\pm=\sqrt{2m(E_F\pm E)-k_y^2}.
\end{equation}
In practic we have assumed chiral p-wave superconducting gaps similar to the A-phase in liquid $^3$He. In this so-called chiral p-wave state where intrinsically the time reversal symmetry is broken,\cite{Ikegami2013,Wu2018} FSs feel different gaps for the hole like quasiparticles and electron like quasiparticles with the energy E. As well, the $\gamma_{+,-}$ factors refer to the momentum direction of the quasiparticles,
\begin{equation}
\hspace*{-0.5cm}e^{\pm i\gamma_{\ell \beta}^{\sigma}}=\frac{\beta k_x^+ \pm i k_y}{\sqrt{k^+=(k_x^+)^2+(k_y)^2}}e^{\pm i\theta^{\ell \sigma}}e^{-i\phi_{\ell \sigma}}, \quad \beta=\pm 1
\end{equation}
\begin{equation}
\hspace*{-0.4cm} e^{\pm i\bar{\gamma}_{\ell \beta}^{\sigma}}=\frac{\beta k_x^- \pm i k_y}{\sqrt{k^-=(k_x^-)^2+(k_y)^2}}e^{\pm i\theta^{\ell \sigma}}e^{-i\phi_{\ell \sigma}}, \quad \beta=\pm 1
\end{equation}
\begin{equation}
\hspace*{-0.4cm}q_{\ell e(h)}^{\sigma}=\sqrt{2m(E_F+\sigma h_{\ell}\pm \sqrt {E^2-|\Delta_{\ell e(h)} {\sigma}|^2})-k_y^2}.
\end{equation}
The wave functions in three regions are written and then the boundary conditions are applied. At the layer interfaces the wavefunction $\psi(r)$ is continuous, while its first derivative has a discontinuity proportional to a dimensionless parameter Z$\equiv$2W$_0$/$\hbar$v$_F$, measuring the height of potential barriers at the interfaces. Here, the junctions without barriers are assumed so Z will be zero. As a result, the four boundary conditions will be inscribed qua
\begin{multline}
\label{boundary}
\hspace*{-0.8cm}\Psi_{FS_1}|_{x=0}=\Psi_{N}|_{x=0},\\
\hspace*{-0.8cm}\Psi_{N}|_{x=L}=\Psi_{FS_2}|_{x=L},\\
\hspace*{-0.8cm}\frac{d\Psi_{FS_1}}{dx}|_{x=0}=\frac{d\Psi_{N}}{dx}|_{x=0},\\
\hspace*{-0.8cm}\frac{d\Psi_{N}}{dx}|_{x=L}=\frac{d\Psi_{FS_2}}{dx}|_{x=L}.
\end{multline}
Therefore, a system of 16 linear equations with many unknown complex parameters are yielded. By solving the matrix, the amplitudes of c$_{j\pm}^{j=1-4}$,r$_{e(h)}^{\sigma}$,t$_{e(h)}^{\sigma}$ will be found. In the following the boundary conditions are applied and the coefficients are computed. For the FS$_1$/N boundary we have
\begin{equation}
\hspace*{-0.8cm}\Psi_{FS_1}|_{x=0}=\Psi_{N}|_{x=0}.
\end{equation}
Thus following four equations are established,
\begin{multline}
\nonumber \hspace*{-0.4cm}
c_{1+}+c_{1-}=(1+r_e^\uparrow) \cos{(\frac{\alpha}{2})}u_{1e}^{\uparrow}+r_e^\downarrow[i\sin{(\frac{\alpha}{2})}u_{1e}^{\downarrow}] \\ +r_h^\uparrow \cos{(\frac{\alpha}{2})}v_{1e}^{\uparrow}e^{i{\bar \gamma}_{1+}^\uparrow}
+\nonumber r_h^\downarrow [i\sin{(\frac{\alpha}{2})}v_{1e}^{\downarrow}e^{i{\bar \gamma}_{1+}^\downarrow}]
\end{multline}
\begin{multline}
\hspace*{-0.4cm}\Longrightarrow -c_{1+}-c_{1-}+r_e^\uparrow \cos{(\frac{\alpha}{2})}u_{1e}^{\uparrow}+r_e^\downarrow[i\sin{(\frac{\alpha}{2})}u_{1e}^{\downarrow}] \\ +r_h^\uparrow \cos{(\frac{\alpha}{2})}v_{1e}^{\uparrow}e^{i{\bar \gamma}_{1+}^\uparrow}+r_h^\downarrow [i\sin{(\frac{\alpha}{2})}v_{1e}^{\downarrow}e^{i{\bar \gamma}_{1+}^\downarrow}]=\\ -\cos{(\frac{\alpha}{2})}u_{1e}^{\uparrow},
\end{multline}
\begin{multline}
\hspace*{-0.2cm} c_{2+}+c_{2-}=(1+r_e^\uparrow) [i\sin{(\frac{\alpha}{2})}u_{1e}^{\uparrow}]+r_e^\downarrow[\cos{(\frac{\alpha}{2})}u_{1e}^{\downarrow}] \\ +r_h^\uparrow [i\sin{(\frac{\alpha}{2})}v_{1e}^{\uparrow}e^{i{\bar \gamma}_{1+}^\uparrow}]+r_h^\downarrow \cos{(\frac{\alpha}{2})}v_{1e}^{\downarrow}e^{i{\bar \gamma}_{1+}^\downarrow}
\end{multline}
\begin{multline}
\hspace*{-0.4cm}\Longrightarrow -c_{2+}-c_{2-}+r_e^\uparrow [i\sin{(\frac{\alpha}{2})}u_{1e}^{\uparrow}]+r_e^\downarrow[\cos{(\frac{\alpha}{2})}u_{1e}^{\downarrow}] \\ +
r_h^\uparrow [i\sin{(\frac{\alpha}{2})}v_{1e}^{\uparrow}e^{i{\bar \gamma}_{1+}^\uparrow}]+r_h^\downarrow \cos{(\frac{\alpha}{2})}v_{1e}^{\downarrow}e^{i{\bar \gamma}_{1+}^\downarrow}=\\ -i\sin{(\frac{\alpha}{2})}u_{1e}^{\uparrow},
\end{multline}
\begin{multline}
\hspace*{-0.4cm} c_{3+}+c_{3-}= \cos{(\frac{\alpha}{2})}v_{1e}^{\uparrow}(e^{-i{\bar \gamma}_{1+}^\uparrow}+e^{-i{\bar \gamma}_{1-}^\uparrow} r_e^\uparrow) \\ -r_e^\downarrow [i\sin{(\frac{\alpha}{2})}v_{1e}^{\downarrow}e^{-i{\bar \gamma}_{1-}^\downarrow}]+r_h^\uparrow \cos{(\frac{\alpha}{2})}u_{1e}^{\uparrow}-r_h^\downarrow [i\sin{(\frac{\alpha}{2})}u_{1e}^{\downarrow}]
\end{multline}
\begin{multline}
\hspace*{-0.4cm}\Longrightarrow -c_{3+}-c_{3-}+r_e^\uparrow\cos{(\frac{\alpha}{2})}v_{1e}^{\uparrow}e^{-i{\bar \gamma}_{1-}^\uparrow}-r_e^\downarrow [i\sin{(\frac{\alpha}{2})}v_{1e}^{\downarrow}\\e^{-i{\bar \gamma}_{1-}^\downarrow}] +r_h^\uparrow \cos{(\frac{\alpha}{2})}u_{1e}^{\uparrow}-
r_h^\downarrow [i\sin{(\frac{\alpha}{2})}u_{1e}^{\downarrow}]=-\cos{(\frac{\alpha}{2})}\\v_{1e}^{\uparrow}e^{-i{\bar \gamma}_{1+}^\uparrow},
\end{multline}
\begin{multline}
\hspace*{-0.4cm} c_{4-}+c_{4+}=[i\sin{(\frac{\alpha}{2})}v_{1e}^{\uparrow}](-e^{-i{\bar \gamma}_{1-}^\uparrow}r_e^\uparrow -e^{-i{\bar \gamma}_{1+}^\uparrow})+r_e^\downarrow \cos{(\frac{\alpha}{2})}\\v_{1e}^{\downarrow}e^{-i{\bar \gamma}_{1-}^\downarrow}
-r_h^\uparrow [i\sin{(\frac{\alpha}{2})}u_{1e}^{\uparrow}]+r_h^\downarrow \cos{(\frac{\alpha}{2})}u_{1e}^{\downarrow}
\end{multline}
\begin{multline}
\hspace*{-0.4cm}\Longrightarrow -c_{4-}-c_{4+}-r_e^\uparrow [i\sin{(\frac{\alpha}{2})}v_{1e}^{\uparrow}e^{-i{\bar \gamma}_{1-}^\uparrow}]+r_e^\downarrow \cos{(\frac{\alpha}{2})}v_{1e}^{\downarrow}e^{-i{\bar \gamma}_{1-}^\downarrow}
\\ -r_h^\uparrow [i\sin{(\frac{\alpha}{2})}u_{1e}^{\uparrow}]+r_h^\downarrow \cos{(\frac{\alpha}{2})}u_{1e}^{\downarrow}=i\sin{(\frac{\alpha}{2})}v_{1e}^{\uparrow}e^{-i{\bar \gamma}_{1+}^\uparrow}.
\end{multline}
Also for the FS$_1$/N junction there exist another boundary condition,
\begin{equation}
\hspace*{-0.8cm}\frac{d\Psi_{FS_1}}{dx}|_{x=0}=\frac{d\Psi_{N}}{dx}|_{x=0}.
\end{equation}
Where,
\begin{equation}
\hspace*{-0.8cm}\frac{d\Psi_{N}}{dx}=
\left (
\begin{array}{rl}
\nonumber ik^+(c_{1+}e^{ik^+x}-c_{1-}e^{-ik^+x})\\
\nonumber ik^+(c_{2+}e^{ik^+x}-c_{2-}e^{-ik^+x})\\
\nonumber ik^-(c_{3-}e^{ik^-x}-c_{3+}e^{-ik^-x})\\
ik^-(c_{4-}e^{ik^-x}-c_{4+}e^{-ik^-x})\\
\end{array}
\right )
\end{equation}
and
\begin{multline}
\hspace*{-0.4cm}\nonumber \frac{d\Psi_{FS_1}}{dx}(1,1)=i\cos{(\frac{\alpha}{2})}(q_{1e}^{\uparrow}u_{1e}^{\uparrow}e^{iq_{1e}^\uparrow x}-r_e^\uparrow q_{1e}^{\uparrow}u_{1e}^{\uparrow}e^{-iq_{1e}^\uparrow x} \\ +r_h^\uparrow q_{1h}^{\uparrow}v_{1e}^{\uparrow}e^{i{\bar \gamma}^\uparrow_{1+}}e^{iq_{1h}^\uparrow x})
-\sin{(\frac{\alpha}{2})}(-r_e^{\downarrow}q_{1e}^{\downarrow}u_{1e}^{\downarrow}e^{-iq_{1e}^\downarrow x} \\ +r_h^\downarrow q_{1h}^{\downarrow}v_{1e}^{\downarrow}e^{iq_{1h}^\downarrow x}e^{i{\bar \gamma}^\downarrow_{1+}}),
\end{multline}
\begin{multline}
\hspace*{-0.4cm} \nonumber \frac{d\Psi_{FS_1}}{dx}(2,1)=-\sin{(\frac{\alpha}{2})}(q_{1e}^{\uparrow}u_{1e}^{\uparrow}e^{iq_{1e}^\uparrow x}-r_e^\uparrow q_{1e}^{\uparrow}u_{1e}^{\uparrow}e^{-iq_{1e}^\uparrow x}\\+r_h^\uparrow q_{1h}^{\uparrow}e^{i{\bar \gamma}^\uparrow_{1+}}e^{iq_{1h}^\uparrow x})+
\nonumber i\cos{(\frac{\alpha}{2})}(-r_e^{\downarrow}q_{1e}^{\downarrow}u_{1e}^{\downarrow}e^{-iq_{1e}^\downarrow x}+\\r_h^\downarrow q_{1h}^{\downarrow}v_{1e}^{\downarrow}e^{iq_{1h}^\downarrow x}e^{i{\bar \gamma}^\downarrow_{1+}}),
\end{multline}
\begin{multline}
\hspace*{-0.4cm}\nonumber \frac{d\Psi_{FS_1}}{dx}(3,1)=i\cos{(\frac{\alpha}{2})}(q_{1e}^{\uparrow}v_{1e}^{\uparrow}e^{iq_{1e}^\uparrow x}e^{-i \gamma^\uparrow_{1+}}-r_e^\uparrow q_{1e}^{\uparrow}v_{1e}^{\uparrow}e^{-iq_{1e}^\uparrow x} \\ e^{-i \gamma^\uparrow_{1-}}+r_h^\uparrow q_{1h}^{\uparrow}u_{1e}^{\uparrow}e^{iq_{1h}^\uparrow x})
\nonumber +\sin{(\frac{\alpha}{2})}(-r_e^{\downarrow}q_{1e}^{\downarrow}v_{1e}^{\downarrow}e^{-iq_{1e}^\downarrow x}e^{-i\gamma^\downarrow_{1-}} \\ +r_h^\downarrow q_{1h}^{\downarrow}u_{1e}^{\downarrow}e^{iq_{1h}^\downarrow x}),
\end{multline}
\begin{multline}
\hspace*{-0.4cm}\nonumber \frac{d\Psi_{FS_1}}{dx}(4,1)=\sin{(\frac{\alpha}{2})}(q_{1e}^{\uparrow}v_{1e}^{\uparrow}e^{iq_{1e}^\uparrow x}e^{-i \gamma^\uparrow_{1+}}-r_e^\uparrow q_{1e}^{\uparrow}v_{1e}^{\uparrow}e^{-iq_{1e}^\uparrow x} \\ e^{-i\gamma^\uparrow_{1-}}+r_h^\uparrow q_{1h}^{\uparrow}u_{1e}^{\uparrow}e^{iq_{1h}^\uparrow x})+
\nonumber i\cos{(\frac{\alpha}{2})}(-r_e^{\downarrow}q_{1e}^{\downarrow}v_{1e}^{\downarrow}e^{-i \gamma^\downarrow_{1-}}e^{-iq_{1e}^\downarrow x} \\ +r_h^\downarrow q_{1h}^{\downarrow}u_{1e}^{\downarrow}e^{iq_{1h}^\downarrow x}).
\end{multline}
The derivatives and the second boundary condition are exploited to gain the next four equations,
\begin{multline}
\hspace*{-0.3cm}
ik^+(-c_{1-}+c_{1+})-(r_e^\uparrow-1)[i\cos{(\frac{\alpha}{2})}q_{1e}^\uparrow u_{1e}^{\uparrow}]+r_h^\uparrow [i\cos{(\frac{\alpha}{2})} \\ q^\uparrow_{1h}v_{1e}^{\uparrow}e^{i{\bar \gamma}^\uparrow_{1+}}]+
r_e^\downarrow \sin{(\frac{\alpha}{2})}q_{1e}^\downarrow u_{1e}^{\downarrow}-r_h^\downarrow\sin{(\frac{\alpha}{2})}q_{1h}^\downarrow v_{1e}^{\downarrow}e^{i{\bar \gamma}^\downarrow_{1+}}]
\end{multline}
\begin{multline}
\hspace*{-0.35cm}
\Longrightarrow ik^+(c_{1-}-c_{1+})-r_e^\uparrow[i\cos{(\frac{\alpha}{2})}q^\uparrow_{1e}u_{1e}^{\uparrow}]+r_h^\uparrow [i\cos{(\frac{\alpha}{2})} \\ q^\uparrow_{1h}v_{1e}^{\uparrow}e^{i{\bar \gamma}^\uparrow_{1+}}]+
r_e^\downarrow \sin{(\frac{\alpha}{2})}q_{1e}^\downarrow u_{1e}^{\downarrow}-r_h^\downarrow\sin{(\frac{\alpha}{2})}q_{1h}^\downarrow v_{1e}^{\downarrow}e^{i{\bar \gamma}^\downarrow_{1+}}] \\ =-i\cos{(\frac{\alpha}{2})}q_{1e}^\uparrow u_{1e}^{\uparrow},
\end{multline}
\begin{multline}
\hspace*{-0.3cm}i k^+(-c_{2-}+c_{2+})+(r_e^\uparrow-1)\sin{(\frac{\alpha}{2})}q^\uparrow_{1e}u_{1e}^{\uparrow}-r_e^\downarrow[i\cos{(\frac{\alpha}{2})} \\ q_{1e}^\downarrow u_{1e}^{\downarrow}]-r_h^\uparrow[\sin{(\frac{\alpha}{2})}q^\uparrow_{1h}e^{i{\bar \gamma}^\uparrow_{1+}}]+
r_h^\downarrow [i\cos{(\frac{\alpha}{2})}q_{1h}^\downarrow v_{1e}^{\downarrow}e^{i{\bar \gamma}^\downarrow_{1+}}]
\end{multline}
\begin{multline}
\hspace*{-0.43cm}
\Longrightarrow i k^+(c_{2-}-c_{2+})+r_e^\uparrow\sin{(\frac{\alpha}{2})}q^\uparrow_{1e}u_{1e}^{\uparrow}-r_e^\downarrow[i\cos{(\frac{\alpha}{2})}q_{1e}^\downarrow u_{1e}^{\downarrow}] \\ -r_h^\uparrow[\sin{(\frac{\alpha}{2})}q^\uparrow_{1h}e^{i{\bar \gamma}^\uparrow_{1+}}]+
r_h^\downarrow [i\cos{(\frac{\alpha}{2})}q_{1h}^\downarrow v_{1e}^{\downarrow}e^{i{\bar \gamma}^\downarrow_{1+}}]=\sin{(\frac{\alpha}{2})} \\ q_{1e}^{\uparrow}u_{1e}^{\uparrow},
\end{multline}
\begin{multline}
\hspace*{-0.3cm} i{k^-}(-c_{3+}+c_{3-})=(e^{-i\gamma^\uparrow_{1+}}-r_e^\uparrow e^{-i\gamma^\uparrow_{1-}}) [i\cos{(\frac{\alpha}{2})}q^\uparrow_{1e}v_{1e}^{\uparrow}] \\ +r_h^\uparrow [i\cos{(\frac{\alpha}{2})}q^\uparrow_{1h}u_{1e}^{\uparrow}]-
r_e^\downarrow \sin{(\frac{\alpha}{2})}q_{1e}^\downarrow v_{1e}^{\downarrow}e^{-i\gamma^\downarrow_{1-}}+r_h^\downarrow\sin{(\frac{\alpha}{2})} \\ q_{1h}^\downarrow u_{1e}^{\downarrow}
\end{multline}
\begin{multline}
\hspace*{-0.43cm}
\Longrightarrow i{k^-}(c_{3+}-c_{3-})-r_e^\uparrow e^{-i\gamma^\uparrow_{1-}} [i\cos{(\frac{\alpha}{2})}q^\uparrow_{1e}v_{1e}^{\uparrow}]+r_h^\uparrow \\ [i\cos{(\frac{\alpha}{2})}q^\uparrow_{1h}u_{1e}^{\uparrow}]-
r_e^\downarrow \sin{(\frac{\alpha}{2})}q_{1e}^\downarrow v_{1e}^{\downarrow}e^{-i\gamma^\downarrow_{1-}}+r_h^\downarrow\sin{(\frac{\alpha}{2})}q_{1h}^\downarrow \\u_{1e}^{\downarrow}=-i\cos{(\frac{\alpha}{2})}q_{1e}^\uparrow v_{1e}^{\uparrow}e^{-i\gamma^\uparrow_{1+}},
\end{multline}
\begin{multline}
\hspace*{-0.3cm}i{k^-}(c_{4+}-c_{4-})-r_e^\uparrow[\sin{(\frac{\alpha}{2})}q^\uparrow_{1e}v_{1e}^{\uparrow}e^{-i\gamma^\uparrow_{1-}}]+
r_h^\uparrow \sin{(\frac{\alpha}{2})}q^\uparrow _{1h} \\ u_{1e}^{\uparrow}-
r_e^\downarrow [i\cos{(\frac{\alpha}{2})}q_{1e}^\downarrow v_{1e}^{\downarrow}e^{-i\gamma^\downarrow_{1-}}]+r_h^\downarrow[i\cos{(\frac{\alpha}{2})}q_{1h}^\downarrow u_{1e}^{\downarrow}]
\end{multline}
\begin{multline}
\hspace*{-0.43cm}
\Longrightarrow
i{k^-}(c_{4+}-c_{4-})-r_e^\uparrow[\sin{(\frac{\alpha}{2})}q^\uparrow_{1e}v_{1e}^{\uparrow}e^{-i\gamma^\uparrow_{1-}}]+r_h^\uparrow\sin{(\frac{\alpha}{2})} \\ q^\uparrow_{1h}u_{1e}^{\uparrow}-
r_e^\downarrow [i\cos{(\frac{\alpha}{2})}q_{1e}^\downarrow v_{1e}^{\downarrow}e^{-i\gamma^\downarrow_{1-}}]+r_h^\downarrow[i\cos{(\frac{\alpha}{2})}q_{1h}^\downarrow u_{1e}^{\downarrow}] \\ =-\sin{(\frac{\alpha}{2})}v_{1e}^{\uparrow}e^{-i\gamma^\uparrow_{1+}}q_{1e}^{\uparrow}.
\end{multline}
Applying the first boundary conditions for the N/FS$_2$ junction,
\begin{equation}
\hspace*{-0.8cm}\Psi_{N}|_{x=L}=\Psi_{FS_2}|_{x=L},
\end{equation}
the following equations are obtained,
\begin{multline}
\hspace*{-0.4cm}c_{1+}e^{ik^+L}+c_{1-}e^{-ik^+L}=t_e^\uparrow u_{2e}^{\uparrow}e^{iq_{2e}^\uparrow L}+t_h^\uparrow v_{2e}^{\uparrow}e^{i{\bar \gamma}_{2-}^\uparrow}e^{-iq^{\uparrow}_{2h}L}\\ \hspace*{-0.5cm} \Longrightarrow -c_{1+}e^{ik^+L}-c_{1-}e^{-ik^+L}+t_e^\uparrow u_{2e}^{\uparrow}e^{iq_{2e}^\uparrow L}+t_h^\uparrow v_{2e}^{\uparrow}e^{i{\bar \gamma}_{2-}^\uparrow}\\e^{-iq^{\uparrow}_{2h}L}=0,
\end{multline}
\begin{multline}
\hspace*{-0.4cm}c_{2+}e^{ik^+L}+c_{2-}e^{-ik^+L}=t_e^\downarrow u_{2e}^{\downarrow}e^{iq_{2e}^\downarrow L}+t_h^\downarrow v_{2e}^{\downarrow}e^{i{\bar \gamma}^{\downarrow}_{2-}}e^{-iq_{2h}^\downarrow L}\\ \hspace*{-0.5cm}\Longrightarrow -c_{2+}e^{ik^+L}-c_{2-}e^{-ik^+L}+t_e^\downarrow u_{2e}^{\downarrow}e^{iq_{2e}^\downarrow L}+t_h^\downarrow v_{2e}^{\downarrow}e^{i{\bar \gamma}^{\downarrow}_{2-}} \\ e^{-iq_{2h}^\downarrow L}=0,
\end{multline}
\begin{multline}
\hspace*{-0.4cm}c_{3-}e^{i{k^-}L}+c_{3+}e^{-i{k^-}L}=t_e^\uparrow v_{2e}^{\uparrow}e^{-i\gamma_{2+}^\uparrow}e^{iq^{\uparrow}_{2e}L}+t_h^\uparrow u_{2e}^{\uparrow}e^{-iq_{2h}^\uparrow L}\\ \hspace*{-0.3cm} \Longrightarrow -c_{3-}e^{i{k^-}L}-c_{3+}e^{-i{k^-}L}+t_e^\uparrow v_{2e}^{\uparrow}e^{-i\gamma_{2+}^\uparrow}e^{iq^{\uparrow}_{2e}L}+t_h^\uparrow u_{2e}^{\uparrow} \\ e^{-iq_{2h}^\uparrow L}=0,
\end{multline}
\begin{multline}
\hspace*{-0.4cm} c_{4-}e^{i{k^-}L}+c_{4+}e^{-i{k^-}L}=t_e^\downarrow v_{2e}^{\downarrow}e^{-i\gamma_{2+}^\downarrow}e^{iq^{\downarrow}_{2e}L}+t_h^\downarrow u_{2e}^{\downarrow}e^{-iq_{2h}^\downarrow L} \\ \hspace*{-0.3cm} \Longrightarrow -c_{4-}e^{i{k^-}L}-c_{4+}e^{-i{k^-}L}+t_e^\downarrow v_{2e}^{\downarrow}e^{-i\gamma_{2+}^\downarrow}e^{iq^{\downarrow}_{2e}L}+t_h^\downarrow u_{2e}^{\downarrow} \\ e^{-iq_{2h}^\downarrow L}=0.
\end{multline}
Also, the second boundary condition at N/FS$_2$ junction is employed to find the last four equations
\begin{multline}
\hspace*{-0.4cm}\frac{d\Psi_{N}}{dx}|_{x=L}=\frac{d\Psi_{FS_2}}{dx}|_{x=L}=
\\
\left (
\begin{array}{rl}
\nonumber i(t^\uparrow_eu_{2e}^{\uparrow}q_{2e}^{\uparrow}e^{iq^\uparrow_{2e}L}-t^\uparrow_h v_{2e}^{\uparrow}q_{2h}^{\uparrow}e^{i{\bar \gamma}^{\uparrow}_{2-}}e^{-iq^\uparrow_{2h}L})\\
\nonumber i(t^\downarrow_eu_{2e}^{\downarrow}q_{2e}^{\downarrow}e^{iq^\downarrow_{2e}L}-t^\downarrow_h v_{2e}^{\downarrow}q_{2h}^{\downarrow}e^{i{\bar \gamma}^{\downarrow}_{2-}}e^{-iq^\downarrow_{2h}L})\\
\nonumber i(t^\uparrow_e v_{2e}^{\uparrow}q_{2e}^{\uparrow}e^{-i{\bar \gamma}^{\uparrow}_{2+}}e^{iq^\uparrow_{2e}L}-t^\uparrow_h u_{2e}^{\uparrow}q_{2h}^{\uparrow}e^{-iq^\uparrow_{2h}L})\\
i(t^\downarrow_e v_{2e}^{\downarrow}q_{2e}^{\downarrow}e^{-i{\bar \gamma}^{\downarrow}_{2+}}e^{iq^\downarrow_{2e}L}-t^\downarrow_Lu_{2e}^{\downarrow}q_{2h}^{\downarrow}e^{-iq^\downarrow_{2h}L})\\
\end{array}
\right ).
\end{multline}
Accordingly following equations are obtained,
\begin{multline}
\hspace*{-0.3cm}ik^+ (-c_{1-}e^{-ik^+L}+c_{1+}e^{ik^+L})=t_e^\uparrow (i u_{2e}^{\uparrow}q_{2e}^\uparrow e^{iq_{2e}^\uparrow L})-t_h^\uparrow(iv_{2e}^{\uparrow} \\ q_{2h}^\uparrow e^{i{\bar \gamma}_{2-}^\uparrow}e^{-iq^\uparrow_{2h}L}) \Longrightarrow ik^+ (c_{1-}e^{-ik^+L}-c_{1+}e^{ik^+L})+t_e^\uparrow \\ (iq_{2e}^\uparrow u_{2e}^{\uparrow}e^{iq_{2e}^\uparrow L})-t_h^\uparrow(iq_{2h}^\uparrow v_{2e}^{\uparrow} e^{i{\bar \gamma}_{2-}^\uparrow} e^{-iq^\uparrow_{2h}L})=0,
\end{multline}
\begin{multline}
\hspace*{-0.3cm} ik^+ (-c_{2-}e^{-ik^+L}+c_{2+}e^{ik^+L})=t_e^\downarrow (i u_{2e}^{\downarrow}q_{2e}^\downarrow e^{iq_{2e}^\downarrow L})-t_h^\downarrow(iv_{2e}^{\downarrow} \\ q_{2h}^\downarrow e^{i{\bar \gamma}_{2-}^\downarrow}e^{-iq^\downarrow_{2h}L}) \Longrightarrow ik^+ (c_{2-}e^{-ik^+L}-c_{2+}e^{ik^+L})+t_e^\downarrow \\ (i u_{2e}^{\downarrow}q_{2e}^\downarrow e^{iq_{2e}^\downarrow L})-t_h^\downarrow(iv_{2e}^{\downarrow}q_{2h}^\downarrow e^{i{\bar \gamma}_{2-}^\downarrow}e^{-iq^\downarrow_{2h}L})=0,
\end{multline}
\begin{multline}
\hspace*{-0.3cm}ik^- (-c_{3+}e^{-ik^-L}+c_{3-}e^{ik^- L})=t_e^\uparrow(iq_{2e}^\uparrow v_{2e}^{\uparrow} e^{-i\gamma_{2+}^\uparrow }e^{iq^{\uparrow}_{2e}L})-t_h^\uparrow \\ (iq_{2h}^\uparrow u_{2e}^{\uparrow} e^{-iq_{2h}^\uparrow L}) \Longrightarrow ik^- (c_{3+}e^{-ik^-L}-c_{3-}e^{ik^- L})+t_e^\uparrow \\ (iq_{2e}^\uparrow v_{2e}^{\uparrow}e^{-i\gamma_{2+}^\uparrow }e^{iq^{\uparrow}_{2e}L})-t_h^\uparrow(iq_{2h}^\uparrow u_{2e}^{\uparrow} e^{-iq_{2h}^{\uparrow} L})=0,
\end{multline}
\begin{multline}
\hspace*{-0.4cm}ik^- (-c_{4+}e^{-ik^- L}+c_{4-}e^{ik^- h})=t_e^\downarrow (iq_{2e}^\downarrow v_{2e}^{\downarrow} e^{-i\gamma_{2+}^\downarrow }e^{iq_{2e}^\downarrow}h)-t_h^\downarrow \\ (iq_{2h}^\downarrow u_{2e}^{\downarrow} e^{-iq_{2h}^\downarrow L}) \Longrightarrow
ik^- (c_{4+}e^{-ik^- L}-c_{4-}e^{ik^- L})+t_e^\downarrow \\ (iq_{2e}^\downarrow v_{2e}^{\downarrow} e^{-i\gamma_{2+}^\downarrow }e^{iq_{2e}^\downarrow}h)-t_h^\downarrow(iq_{2h}^\downarrow u_{2e}^{\downarrow} e^{-iq_{2h}^\downarrow L})=0.
\end{multline}
So the subsequent formula holds,
\begin{equation}
\hspace*{-0.8cm}\textbf{Ab=C}.
\end{equation}
\newline
Where,
\begin{multline}
\hspace*{-0.4cm}
b^T=( c_{1+} \ c_{1-} \ c_{2+} \ c_{2-} \ c_{3+} \ c_{3-} \ c_{4+} \ c_{4-} \ r_e^\uparrow \ r_e^\downarrow \ r_h^\uparrow \ r_h^\downarrow \ t_e^\uparrow \ \\ t_e^\downarrow \ t_h^\uparrow \ t_h^\downarrow)\end{multline}
and
\begin{equation}
\label{4}
\hspace*{-0.8cm} \textbf{A}=\left(
\begin{array}{cc}
\check{A}_{11} & \check{A}_{12} \\
\check{A}_{21} & \check{A}_{22} \\
\end{array}
\right)
\end{equation}
\begin{equation}
\hspace*{-0.4cm}\nonumber \check{A}_{11}=\left(
\begin{array}{cccccccc}
-1 & -1 & 0 & 0 & 0 & 0 & 0 & 0 \\
0 & 0 & -1 & -1 & 0 & 0 & 0 & 0 \\
0 & 0 & 0 & 0 & -1 & -1 & 0 & 0 \\
0 & 0 & 0 & 0 & 0 & 0 & -1 & -1 \\
-1 & 1 & 0 & 0 & 0 & 0 & 0 & 0 \\
0 & 0 & -1 & 1 & 0 & 0 & 0 & 0 \\
0 & 0 & 0 & 0 & 1 & -1 & 0 & 0 \\
0 & 0 & 0 & 0 & 0 & 0 & 1 & -1 \\
\end{array}
\right),
\end{equation}
\begin{widetext}
$\hspace*{-0.0cm}\nonumber \check{A}_{12}=\left(
\begin{array}{cccccccc}
\cos{\frac{\alpha}{2}}u_{1e}^{\uparrow} & i\sin{\frac{\alpha}{2}}u_{1e}^{\downarrow} & \cos{\frac{\alpha}{2}}v_{1e}^{\uparrow}e^{i{\bar \gamma}_{1+}^{\uparrow}} & i\sin{\frac{\alpha}{2}}v_{1e}^{\downarrow}e^{i{\bar \gamma}_{1+}^{\downarrow}} & 0 & 0 &0 & 0 \\
i\sin{\frac{\alpha}{2}}u_{1e}^{\uparrow} & \cos{\frac{\alpha}{2}}u_{1e}^{\downarrow} & i\sin{\frac{\alpha}{2}}v_{1e}^{\uparrow}e^{i{\bar \gamma}_{1+}^{\uparrow}} & \cos{\frac{\alpha}{2}}v_{1e}^{\downarrow}e^{i{\bar \gamma}_{1+}^{\downarrow}} & 0 & 0 & 0 & 0 \\
\cos{\frac{\alpha}{2}}v_{1e}^{\uparrow}e^{-i \gamma_{1-}^{\uparrow}} & -i\sin{\frac{\alpha}{2}}v_{1e}^{\downarrow}e^{-i \gamma_{1-}^{\downarrow}} & \cos{\frac{\alpha}{2}}u_{1e}^{\uparrow} & -i\sin{\frac{\alpha}{2}}u_{1e}^{\downarrow} & 0 & 0 & 0 & 0 \\
-i\sin{\frac{\alpha}{2}}v_{1e}^{\uparrow}e^{-i \gamma_{1-}^{\uparrow}} & \cos{\frac{\alpha}{2}}v_{1e}^{\downarrow}e^{-i \gamma_{1-}^{\downarrow}} & -i\sin{\frac{\alpha}{2}}u_{1e}^{\uparrow} & \cos{\frac{\alpha}{2}}u_{1e}^{\downarrow} & 0 & 0 & 0 & 0 \\
-\cos{\frac{\alpha}{2}}\frac{q_{1e}^{\uparrow}}{k_x^+}u_{1e}^{\uparrow} & -i\sin{\frac{\alpha}{2}}\frac{q_{1e}^{\downarrow}}{k_x^+}u_{1e}^{\downarrow} & \cos{\frac{\alpha}{2}}\frac{q_{1h}^{\uparrow}}{k_x^+}v_{1e}^{\uparrow}e^{i{\bar \gamma}_{1+}^{\uparrow}} & i\sin{\frac{\alpha}{2}}\frac{q_{1h}^{\downarrow}}{k_x^+}v_{1e}^{\downarrow}e^{i{\bar \gamma}_{1+}^{\downarrow}} & 0 & 0 & 0 & 0 \\
-i\sin{\frac{\alpha}{2}}\frac{q_{1e}^{\uparrow}}{k_x^+}u_{1e}^{\uparrow} & -\cos{\frac{\alpha}{2}}\frac{q_{1e}^{\downarrow}}{k_x^+}u_{1e}^{\downarrow} & i\sin{\frac{\alpha}{2}}\frac{q_{1h}^{\uparrow}}{k_x^+}v_{1e}^{\uparrow}e^{i{\bar \gamma}_{1+}^{\uparrow}} & \cos{\frac{\alpha}{2}}\frac{q_{1h}^{\downarrow}}{k_x^+}v_{1e}^{\downarrow}e^{i{\bar \gamma}_{1+}^{\downarrow}} & 0 & 0 & 0 & 0 \\
-\cos{\frac{\alpha}{2}}\frac{q_{1e}^{\uparrow}}{k_x^-}v_{1e}^{\uparrow}e^{-i\gamma_{1-}^{\uparrow}}& i\sin{\frac{\alpha}{2}}\frac{q_{1e}^{\downarrow}}{k_x^-}v_{1e}^{\downarrow}e^{-i\gamma_{1-}^{\downarrow}} & \cos{\frac{\alpha}{2}}\frac{q_{1h}^{\uparrow}}{k_x^-}u_{1e}^{\uparrow} & -i\sin{\frac{\alpha}{2}}\frac{q_{1h}^{\downarrow}}{k_x^-}u_{1e}^{\downarrow} & 0 & 0 & 0 & 0 \\
i\sin{\frac{\alpha}{2}}\frac{q_{1e}^{\uparrow}}{k_x^-}v_{1e}^{\uparrow}e^{-i\gamma_{1-}^{\uparrow}} & -\cos{\frac{\alpha}{2}}\frac{q_{1e}^{\downarrow}}{k_x^-}v_{1e}^{\downarrow}e^{-i\gamma_{1-}^{\downarrow}} & -i\sin{\frac{\alpha}{2}}\frac{q_{1h}^{\uparrow}}{k_x^-}u_{1e}^{\uparrow} & \cos{\frac{\alpha}{2}}\frac{q_{1h}^{\downarrow}}{k_x^-}u_{1e}^{\downarrow} & 0 & 0 & 0 & 0 \\
\end{array}
\right)$,
\\
\\
\\
\hspace*{0.3cm}\nonumber
$\check{A}_{21}=\left(
\begin{array}{cccccccc}
-e^{ik^{+}L} & -e^{-ik^{+}L} & 0 & 0 & 0 & 0 & 0 & 0 \\
0 & 0 & -e^{ik^{+}L} & -e^{-ik^{+}L} & 0 & 0 & 0 & 0 \\
0 & 0 & 0 & 0 & -e^{-ik^{-}L} & -e^{ik^{-}L} & 0 & 0 \\
0 & 0 & 0 & 0 & 0 & 0 & -e^{-ik^{-}L} & -e^{ik^{-}L} \\
-e^{ik^{+}L} & e^{-ik^{+}L} & 0 & 0 & 0 & 0 & 0 & 0 \\
0 & 0 & -e^{ik^{+}L} & e^{-ik^{+}L} & 0 & 0 & 0 & 0 \\
0 & 0 & 0 & 0 & e^{-ik^{-}L} & -e^{ik^{-}L} & 0 & 0 \\
0 & 0 & 0 & 0 & 0 & 0 & e^{-ik^{-}L} & -e^{ik^{-}L} \\
\end{array}
\right)$,
\nonumber
\\
\hspace*{0.3cm}\nonumber
$\check{A}_{22}=\left(
\begin{array}{cccccccc}
0 & 0 & 0 & 0 & u_{2e}^{\uparrow}e^{iq_{2e}^{\uparrow}L} & 0 & v_{2e}^{\uparrow}e^{i{\bar \gamma}_{2-}^{\uparrow}}e^{-iq_{2h}^{\uparrow}L} & 0 \\
0 & 0 & 0 & 0 & 0 & u_{2e}^{\downarrow}e^{iq_{2e}^{\downarrow}L} & 0 & v_{2e}^{\downarrow}e^{i{\bar \gamma}_{2-}^{\downarrow}}e^{-iq_{2h}^{\downarrow}L} \\
0 & 0 & 0 & 0 & v_{2e}^{\uparrow}e^{-i\gamma_{2+}^{\uparrow}}e^{iq_{2e}^{\uparrow}L} & 0 & u_{2e}^{\uparrow}e^{-iq_{2h}^{\uparrow}L} & 0 \\
0 & 0 & 0 & 0 & 0 & v_{2e}^{\downarrow}e^{-i\gamma_{2+}^{\downarrow}}e^{iq_{2e}^{\downarrow}L} & 0 & u_{2e}^{\downarrow}e^{-iq_{2h}^{\downarrow}L} \\
0 & 0 & 0 & 0 & \frac{q_{2e}^{\uparrow}}{k_x^+}u_{2e}^{\uparrow}e^{iq_{2e}^{\uparrow}L} & 0 & -\frac{q_{2h}^{\uparrow}}{k_x^+} v_{2e}^{\uparrow}e^{i\gamma_{2-}^{\uparrow}}e^{-iq_{2h}^{\uparrow}L} & 0 \\
0 & 0 & 0 & 0 & 0 & \frac{q_{2e}^{\downarrow}}{k_x^+}u_{2e}^{\downarrow}e^{iq_{2e}^{\downarrow}L} & 0 & -\frac{q_{2h}^{\downarrow}}{k_x^+} v_{2e}^{\downarrow}e^{i{\bar \gamma}_{2-}^{\downarrow}}e^{-iq_{2h}^{\downarrow}L} \\
0 & 0 & 0 & 0 & \frac{q_{2e}^{\uparrow}}{k_x^-} v_{2e}^{\uparrow}e^{-i\gamma_{2+}^{\uparrow}}e^{iq_{2e}^{\uparrow}L} & 0 & -\frac{q_{2h}^{\uparrow}}{k_x^-}u_{2e}^{\uparrow}e^{-iq_{2h}^{\uparrow}L} & 0 \\
0 & 0 & 0 & 0 & 0 & \frac{q_{2e}^{\downarrow}}{k_x^-} v_{2e}^{\downarrow}e^{-i\gamma_{2+}^{\downarrow}}e^{iq_{2e}^{\downarrow}L} & 0 & -\frac{q_{2h}^{\downarrow}}{k_x^-}u_{2e}^{\downarrow}e^{-iq_{2h}^{\downarrow}L}, \\
\end{array}
\right)$.
\end{widetext}
and more the matrix C is
\begin{widetext}
$C^{T}=(-\cos{(\frac{\alpha}{2})}u_{1e}^{\uparrow}\quad-i\sin{(\frac{\alpha}{2})}u_{1e}^{\uparrow}\quad-\cos{(\frac{\alpha}{2})}v_{1e}^{\uparrow}e^{-i{\bar \gamma}_{1+}^\uparrow}\quad
i\sin{(\frac{\alpha}{2})}v_{1e}^{\uparrow}e^{-i{\bar \gamma}_{1+}^\uparrow}
-i\cos{(\frac{\alpha}{2})}q_{1e}^\uparrow u_{1e}^{\uparrow}\quad \sin{(\frac{\alpha}{2})}q_{1e}^{\uparrow}u_{1e}^{\uparrow}\quad
-i\cos{(\frac{\alpha}{2})}q_{1e}^\uparrow v_{1e}^{\uparrow}e^{-i\gamma^\uparrow_{1+}}\quad -\sin{(\frac{\alpha}{2})}v_{1e}^{\uparrow}e^{-i\gamma^\uparrow_{1+}}q_{1e}^{\uparrow}
0\quad\quad\quad\quad 0\quad\quad\quad\quad 0\quad\quad\quad\quad 0\quad\quad\quad\quad 0\quad\quad\quad\quad 0\quad\quad\quad\quad 0\quad\quad\quad\quad 0 )$.
\end{widetext}
This linear system should be solved to find the unknown coefficients of matrix b. As soon as the linear system is solved and the coefficients are known, they are used to find the STT in each FSs via the following formula
\begin{equation}
\label{tausst2}
\hspace*{-0.3cm}\tau_{stt}=\psi^{\dag}\left(
\begin{array}{cc}
\sigma\times \textbf{h} & 0 \\
0 & -\sigma^{*}\times \textbf{h} \\
\end{array}
\right)
\psi.
\end{equation}
\begin{equation}
\hspace*{-0.8cm} \left(
\begin{array}{cc}
\vec{\sigma} & 0 \\
0 & \vec{\sigma}^* \\
\end{array}
\right)=\left(
\begin{array}{cccc}
\hat{z} & \hat{x}-i \hat{y} & 0 & 0 \\
\hat{x}+i \hat{y} & -\hat{z} & 0 & 0 \\
0 & 0 & \hat{z} & \hat{x}+i \hat{y} \\
0 & 0 & \hat{x}-i \hat{y} & -\hat{z} \\
\end{array}
\right)
\end{equation}

Here, the interval of E$<$$\Delta$ is contemplated. With consideration of this limit we proceed to find the STT. The respective contribution of each energy to the spectral currents, spin supercurrent and STT are found and then we continue to the higher energy until the required precision of the summation over energy is achieved.\\
\section{Algorithm}
The first part is involved with calculating the elements of matrix b. The wave function for an electron with specified energy E is written and the lapack library is applied to the 16$\times$16 general complex matrix to calculate the undetermined coefficients of the linear system. Notably the ZGETRF, ZGETRI, ZGESVD, and ZGEEV routines \cite{Lapack1999} call in our code. The input parameters for the boundary conditions and the numerical method are available at \onlinecite{Zahracode3}, which is written in Fortran. Finding matrix b, all the undetermined coefficients in the wave functions become known and the bound state energies can be established. Secondly, the STT is calculated in three directions for the each specified obtained bound state energy E and then is summed to get the final STT. 

\end{document}